\documentclass{aa}
\newcommand{\bit}{\begin{itemize}}
\newcommand{\eit}{\end{itemize}}
\newcommand{\ben}{\begin{enumerate}}
\newcommand{\een}{\end{enumerate}}
\newcommand{\bde}{\begin{description}}
\newcommand{\ede}{\end{description}}

\usepackage{graphicx}
\usepackage{lscape}

\usepackage{natbib}
\bibpunct{(}{)}{;}{a}{}{,}

\begin{document}
\title{Star formation rates and mass distributions in interacting galaxies}
\author{W. Kapferer,
        A. Knapp,
        S. Schindler,
        S. Kimeswenger \&
        E. van Kampen}

\institute{ Institut f\"ur Astrophysik,
            Leopold-Franzens-Universit\"at Innsbruck,
            Technikerstr. 25,
            A-6020 Innsbruck, Austria
}
\offprints{W. Kapferer, \email{wolfgang.e.kapferer@uibk.ac.at}}

\date{-/-}

\abstract{We present a systematic investigation of the star
formation rate (hereafter SFR) in interacting disk galaxies. We
determine the dependence of the overall SFR on different spatial
alignments and impact parameters of more than 50 different
configurations in combined N-body/hydrodynamic simulations. We
also show mass profiles of the baryonic components. We find that
galaxy-galaxy interactions can enrich the surrounding intergalatic
medium with metals very efficiently up to distances of several 100
kpc. This enrichment can be explained in terms of indirect
processes like thermal driven galactic winds or direct processes
like 'kinetic' spreading of baryonic matter. In the case of equal
mass mergers the direct -kinetic- redistribution of gaseous matter
(after 5 Gyr) is less efficient than the environmental enrichment
of the same isolated galaxies by a galactic wind. In the case of
non-equal mass mergers however, the direct -kinetic- process
dominates the redistribution of gaseous matter. Compared to the
isolated systems, the integrated star formation rates (ISFRs)
($\int_{t\,=\,0\,Gyr}^{t\,=\,5\,Gyr}\textnormal{SFR(t)}dt$) in the
modelled interacting galaxies are in extreme cases a factor of 5
higher and on average a factor of 2 higher in interacting
galaxies. Co-rotating and counter-rotating interactions do not
show a common trend for the enhancement of the ISFRs depending on
the interaction being edge-on or face-on. The latter case shows an
increase of the ISFRs for the counter-rotating system of about
100\%, whereas the edge-on counter-rotating case results in a
lower increase ($\sim$ 10\%). An increase in the minimum
separation yields only a very small decrease in the ISFR after the
first encounter. If the minimum separation is larger than $\sim 5
\times$ the disk scale length R$_{d}$ the second encounter does
not provide an enhancement for the ISFR.

 }
\authorrunning {W. Kapferer et al.}
\titlerunning {Star formation rates and mass distributions in interacting galaxies}
\maketitle

%

\section{Introduction}

Optical, far-infrared and radio observations in the last decades
have shown that the global star formation rate (SFR)
[$M_{\sun}$/yr] in interacting disk galaxies is enhanced in
comparison to isolated galaxies (Bushouse H. A. 1987; Sulentic
1976; Stocke 1978). Modern imaging surveys like GEMS (Rix et al.
2004) or COMBO17 (Bell et al. 2004) reveal the importance of
mergers on the evolution of red-sequence/early-type galaxies and
constrain therefore hierarchical models of galaxy formation and
evolution. Studying single objects like elliptical galaxies with
dust and gas layers like NGC6255 (Morganti et al. 2000) shows us
the complexity of merged systems very impressively. Kauffmann et
al. 2004 concluded that the majority of massive red galaxies are
the result of mergers, in which rapidly stars are formed and gas
is depleted. To study the dynamics and evolution of stellar
populations of merging
systems observations as well as simulations are necessary.\\
Whereas former numerical investigations had a special emphasis on
modelling observed interacting systems, like NGC7252 (Mihos et al.
1998), we are interested in the dependence of the star formation
rates on interaction parameters like spatial alignment and minimum
separation. Cox et al. (2004) investigated galaxy mergers with a
special emphasis on the heating process of gas due to shocks.
Simulations including accretion onto supermassive black holes in
merging galaxies (Springel et al. 2005) and the resulting
suppression of star formation and the morphology of the elliptical
remnant are the newest improvements on this topic. It was recently
investigated by Springel \& Hernquist (2005) wether galaxy mergers
always end up in an elliptical galaxy or not. They have shown that
under certain circumstances (e.g. gas rich disk) the merger
remnant can be a star forming disk galaxy. As the global star
formation rate will increase the overall supernova rate (SNR)
[SN/yr] in close pairs of galaxies this increases the mass loss
rates (MLR) [$M_{\sun}$/yr] of such systems due to supernova
driven mass loaded galactic winds (Colina, Lipari \& Macchetto
1991). Modern X-ray astronomy has revealed the non-primordial
metallicity of the intra-cluster medium (ICM) (Tamura et al.
2004). In addition metal maps of galaxy clusters show that the
metals are not uniformly distributed over the ICM (Schmidt et al.
2002; Furusho et al. 2003; Sanders et al. 2004; Fukazawa et al.
2004). As heavy elements are produced in stars the processed
material must have been ejected by cluster galaxies into the ICM.
Ram-pressure stripping (Gunn \& Gott 1972), galactic winds (De
Young 1978) and direct enrichment by galaxy-galaxy interactions
(Gnedin 1998) present possible transport mechanisms. De Lucia et
al. (2004) used combined N-body and semi-analytical techniques to
model the intergalactic and intra-cluster chemical enrichment due
to galactic winds. As mergers can cause superwinds due to enhanced
star formation they play an important role for enrichment
processes. Tornatore et al. (2004) did Tree+SPH simulations of
galaxy clusters to study the metal enrichment of the intra-cluster
medium. All these different approaches need a proper treatment of
galaxy mergers. In this paper we present a detailed study on the
dependence of the SFRs of interacting galaxies on the spatial
orientation and the impact parameter. We show that not only
galactic winds, due to an enhanced star formation, can enrich the
intergalactic or intracluster medium, but direct redistribution
due to the interaction process has to be taken into account as
well. In order to investigate the duration and strength of the
SFRs in interacting galaxies we did SPH simulations with an
updated version of GADGET (Springel et al. 2001), which employs
the entropy conservative formulation (Springel \& Hernquist 2002).
Spiral galaxies with different spatial alignments were thereby put
on different collision trajectories.

\section{The galaxy model}

The simulated galaxies were modelled with an initial condition
generator for disk galaxies developed by Volker Springel. A
detailed description and analysis of the method and the influence
of the initial conditions on the evolution of the model galaxies
can be found in Springel et al. (2004). The mass and the virial
radius of the halo are given by
\begin{equation}
M_{200}=\frac{v^{3}_{200}}{10\,G\,H(z)} \qquad\mbox{and}\qquad
r_{200}=\frac{v_{200}}{10\,H(z)}
\end{equation}
with $H(z)$ being the Hubble constant at redshift z and $G$ the
gravitational constant. We built three different spiral galaxies,
two of them with a bulge and one without a bulge. See Table
\ref{galaxyproperties} for the properties of the model galaxies.
To allow for resolution studies each model galaxy was investigated
with different particle numbers, see Table
\ref{galaxyproperties_resolution}. Figure \ref{Model galaxies}
shows images of the model galaxies at 0.5 Gyr after the start of
the simulation. The distribution of both the gas and the
collisionless particles in the disk are shown edge on as well as
face on. The star forming knots in the disk are visible.

\begin{table*}
\begin{center}
\caption[]{Properties of the model galaxies}
\begin{tabular}{l l l l }
\hline \hline Properties & Galaxy A & Galaxy B & Galaxy C \cr
\hline halo concentration$^{1}$ & 5 & 5 & 5  \cr circular velocity
$v_{200}$ [km/s] $^{2}$& 160 & 80 & 160  \cr spin parameter
$\lambda$ $^{3}$ & 0.05 & 0.05 & 0.05\cr disk mass fraction$^{4}$
& 0.05 & 0.05 & 0.05 \cr bulge mass fraction$^{4}$ & 0 & 0 & 0.025
\cr bulge size$^{5}$ & 0 & 0 & 0.5 \cr gas content in the
disk$^{6}$ & 0.25 & 0.25 & 0.25  \cr disk thickness$^{7}$ & 0.02 &
0.02 & 0.02  \cr HI gas mass fraction$^{8}$ & 0.5 & 0.5 & 0.5 \cr
total mass [$M_{\sun}$] & 1.3375x10$^{11}$ $h^{-1}$ &
1.67188x10$^{10}$ $h^{-1}$& 1.3375x10$^{11}$ $h^{-1}$ \cr disk
scale length [kpc] & 4.50622$h^{-1}$ & 2.25311$h^{-1}$ &
3.91342$h^{-1}$\cr \hline
\end{tabular}
\label{galaxyproperties}
\end{center}
$^{1}$...$c\equiv\frac{r_{200}}{r_{s}}$ ($r_{s}$..scale radius for
dark matter halo density profile
$\rho(r)=\rho_{crit}\frac{\delta_{0}}{(r/r_{s})(1+r/r_{s})^{2}}$)\newline
$^{2}$...circular velocity at $r_{200}$\newline
$^{3}$...$\lambda=J|E|^{1/2}G^{-1}M^{-5/2}$\newline $^{4}$...of
the total mass (baryonic and non baryonic matter)\newline
$^{5}$...bulge scale length in units of disk scale length \newline
$^{6}$...relative content of gas in the disk
\newline $^{7}$...thickness of the disk in units of radial scale
length
\newline $^{8}$...in comparison to the total gas mass
\end{table*}

\begin{table*}
\begin{center}
\caption[]{Particle numbers and mass resolution }
\begin{tabular}{c r r r }
\hline \hline  & Resolution H & Resolution M & Resolution L\cr
\hline Halo & 50000 & 30000 & 10000 \cr Disk collisionless & 35000
& 20000 & 7000 \cr Gas in disk & 35000 & 20000 & 7000 \cr \hline
Total number & 120000 & 70000 & 24000 \cr \hline Mass resolutions
- softening length &  & & \cr [$h^{-1}$ $M_{\sun}$/particle] /
[$h^{-1}$ kpc] & & & \cr Galaxy A Halo & $2.5\times10^{6}$ - 0.2 &
$4.2\times10^{6}$ - 0.4 & $1.3\times10^{7}$ - 0.6 \cr Galaxy A
Disk & $1.9\times10^{5}$ - 0.05 & $3.3\times10^{5}$ - 0.1 &
$9.5\times10^{5}$ - 0.2 \cr Galaxy A Gas & $4.7\times10^{4}$ -
0.05 & $8.4\times10^{4}$ - 0.1 & $2.4\times10^{5}$ - 0.2 \cr
Galaxy B Halo& $3.2\times10^{5}$ - 0.2 & $5.3\times10^{5}$ - 0.4 &
$1.6\times10^{6}$ - 0.6 \cr Galaxy B Disk & $2.4\times10^{4}$ -
0.05 & $4.2\times10^{4}$ - 0.1 & $1.2\times10^{5}$ - 0.2 \cr
Galaxy B Gas& $6.0\times10^{3}$ - 0.05 & $1.1\times10^{4}$ - 0.1 &
$3.0\times10^{4}$ 0.2 \cr \hline Halo & 50000 & 30000 & 10000 \cr
Disk collisionless & 35000 & 20000 & 7000 \cr Gas in disk & 35000
& 20000 & 7000 \cr Bulge particles & 20000 & 10000 & 3500\cr
\hline Total number & 140000 & 80000 & 27500 \cr \hline Mass
resolutions / softening length &  & & \cr [$M_{\sun}$/particle] /
[$h^{-1}$ kpc] & & & \cr Galaxy C Halo & $3.5\times10^{6}$ - 0.2 &
$6.2\times10^{6}$ - 0.4 & $1.8\times10^{7}$ - 0.6 \cr Galaxy C
Disk & $1.9\times10^{5}$ - 0.05 & $3.3\times10^{5}$ - 0.1 &
$9.5\times10^{5}$ - 0.2 \cr Galaxy C Bulge & $1.7\times10^{5}$ -
0.05 & $3.3\times10^{5}$ - 0.1 & $9.5\times10^{5}$ - 0.2 \cr
Galaxy C Gas & $4.7\times10^{4}$ - 0.05 & $8.4\times10^{4}$ - 0.1
& $2.4\times10^{5}$ - 0.2 \cr \hline
\end{tabular}
\label{galaxyproperties_resolution}
\end{center}
\end{table*}

\begin{figure*}
\begin{center}
\includegraphics[width=18cm]{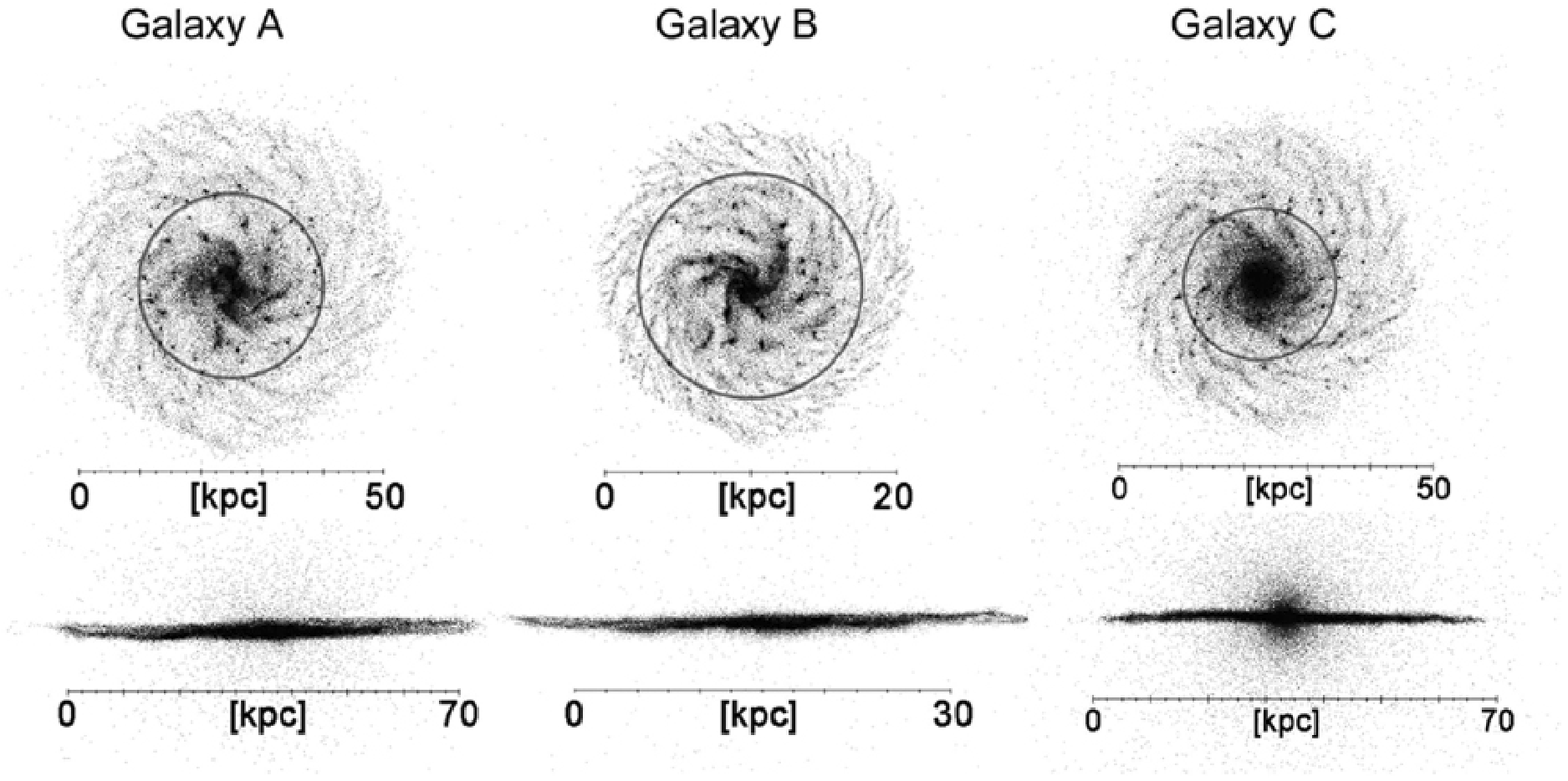}
\caption{Model galaxies A, B and C. The properties of the galaxies
are listed in Table \ref{galaxyproperties}. The upper panels show
gas and collisionless particles in the disk seen face on and the
lower panel shows the galaxies edge on. The ring in the upper
panel marks the optical radius $R_{opt}=3.2R_{d}$ ($R_{d}$...disk
scale length). The knots in the disk represent star forming
regions resulting from gas overdensities. All galaxies are shown
at the same time (~0.5 Gyr after simulation start). The images
correspond to the model galaxies with the highest resolution as
given in Table \ref{galaxyproperties_resolution}2.} \label{Model
galaxies}
\end{center}
\end{figure*}

\section{The star formation and feedback model}

To simulate the feedback of supernovae (SNe) on the interstellar
medium (ISM) we applied the so called "hybrid" method for star
formation and feedback which was introduced by Springel \&
Hernquist (2003). In this hybrid approach condensed cold gas
clouds coexist in pressure equilibrium with a hot ambient gas.
Labelling the average density of the stars as $\rho_{*}$, the
density of the cold gas $\rho_{c}$ and the density of the hot
medium $\rho_{h}$, the total gas density in the disk can be
written as $\rho=\rho_{h}+\rho_{c}$. Because of the finite number
of particles in our simulations $\rho_{*}$ and $\rho$ represent
averages over regions of the inter-stellar medium (ISM). The
central assumption in this approach is the conversion of cold
clouds into stars on a characteristic timescale $t_{*}$ and the
release of a certain mass fraction $\beta$ due to supernovae
(SNe). This relation can be expressed as

\begin{equation}
\frac{d\rho_{*}}{dt}=\frac{\rho_{c}}{t_{*}}-\beta\frac{\rho_{c}}{t_{*}}=(1-\beta)\frac{\rho_{c}}{t_{*}}.
\end{equation}
As a consequence of the fact that released material from SNe is
hot gas the cold gas reservoir increases at the rate
$\rho_{c}/t_{*}$. In this picture $\beta$ gives the mass fraction
of massive stars ($>$ 8 $M_{\sun}$). As in Springel \& Hernquist
(2003) we adopt a $\beta=0.1$ assuming a Salpeter IMF with a slope
of -1.35 in the limits of 0.1 $M_{\sun}$ and 40 $M_{\sun}$. In
addition each supernova event releases energy of $1\times10^{51}$
erg into the ambient medium. This leads to an average return of
$\epsilon=4\times10^{48}$ erg $M_{\sun}^{-1}$. This energy heats
the ambient hot medium and evaporates the cold clouds inside the
hot bubbles of exploding SNe. The total mass of clouds evaporated
can be written as

\begin{equation}
\frac{d\rho_{c}}{dt}|_{EV}=- A\beta\frac{\rho_{c}}{t_{*}}.
\end{equation}
The evaporation process is supposed to be a function of the local
environment with an efficiency A $\propto\rho^{-4/5}$ (McKee \&
Ostriker, 1977). The minimum temperature the gas can reach due to
radiative cooling is about $10^{4}$ K.  The energy balance per
unit volume is $\epsilon_{UV}=\rho_{h}*u_{h}+\rho_{c}*u_{c}$
($u_{h,c}$ energies per unit mass of the cold and hot gas). All
this assumptions lead to self regulating star formation due to the
conversion of cold gas into stars and the feedback of hot gas into
the reservoir of the hot ambient medium, which can cool due to
radiative cooling to cold clouds. The rates are finally

\begin{equation}
\frac{d\rho_{c}}{dt}=-\frac{\rho_c}{t_{*}}-A\beta\frac{\rho_{c}}{t_{*}}+\frac{1-f}{u_{h}-u_{c}}\Lambda_{net}(\rho_{h},u_{h})\\
\end{equation}
\begin{equation}
\frac{d\rho_{h}}{dt}=\beta\frac{\rho_c}{t_{*}}+A\beta\frac{\rho_{c}}{t_{*}}-\frac{1-f}{u_{h}-u_{c}}\Lambda_{net}(\rho_{h},u_{h})
\end{equation}
where f represents a factor to differentiate between ordinary
cooling (f=1) and thermal instability (f=0). $\Lambda_{net}$ is
the cooling function (Katz et al., 1996). Equations (4) and (5)
give the mass budget for the hot and cold gas due to star
formation, mass feedback, cloud evaporation and growth of clouds
due to radiative cooling. On the other side the thermal budget can
be written as

\begin{equation}
\frac{d}{dt}(\rho_{c}u_{c})=-\frac{\rho_c}{t_{*}}u_{c}-A\beta\frac{\rho_{c}}{t_{*}}u_{c}+\frac{(1-f)u_{c}}{u_{h}-u_{c}}\Lambda_{net}
\end{equation}
\begin{equation}
\frac{d}{dt}(\rho_{h}u_{h})=\beta\frac{\rho_c}{t_{*}}(u_{SN}+u_{c})+A\beta\frac{\rho_{c}}{t_{*}}u_{c}-\frac{u_{h}-fu_{c}}{u_{h}-u_{c}}\Lambda_{net}
\end{equation}
with $u_{SN}$ the energy per unit volume from SN explosions. A
major assumption is the fixed temperature of cold gas clouds which
leads to a constant $u_{c}$. One of the basic equations of the
routine is the evolution of the hot phase according to

\begin{equation}
\rho_{h}\frac{du_{h}}{dt}=\beta\frac{\rho_{c}}{t_{*}}(u_{SN}+u_{c}-u_{h})-A\beta\frac{\rho_{c}}{t_{*}}(u_{h}-u_{c})-f\Lambda_{net}.
\end{equation}
From observations and other simulations it is evident that a
certain fraction of matter can escape the galaxies potential due
to thermal and or cosmic ray driven winds due to SNe explosions
(Breitschwerdt et al. 1991). This of course leads to a deficit in
the mass and energy budget of a galaxy, especially in the case of
starbursts. Therefore we applied the same approach as Springel \&
Hernquist (2003) to model this process. The mass-loss rate of the
disk $\dot{M_{w}}$ is proportional to the star formation rate
$\dot{M_{*}}$, in particular $\dot{M_{w}}=\eta\dot{M_{*}}$ with
$\eta=2$, consistent with the observations of Martin (1999). An
additional assumption is that the wind contains a fixed fraction
$\chi$ of SN energy. This fixed fraction is assumed to be 0.25.
For a more detailed description of the feedback routine see
Springel \& Hernquist (2003).

\section{Resolution study for isolated galaxies}

\begin{figure*}
\centering
\includegraphics[width=\textwidth]{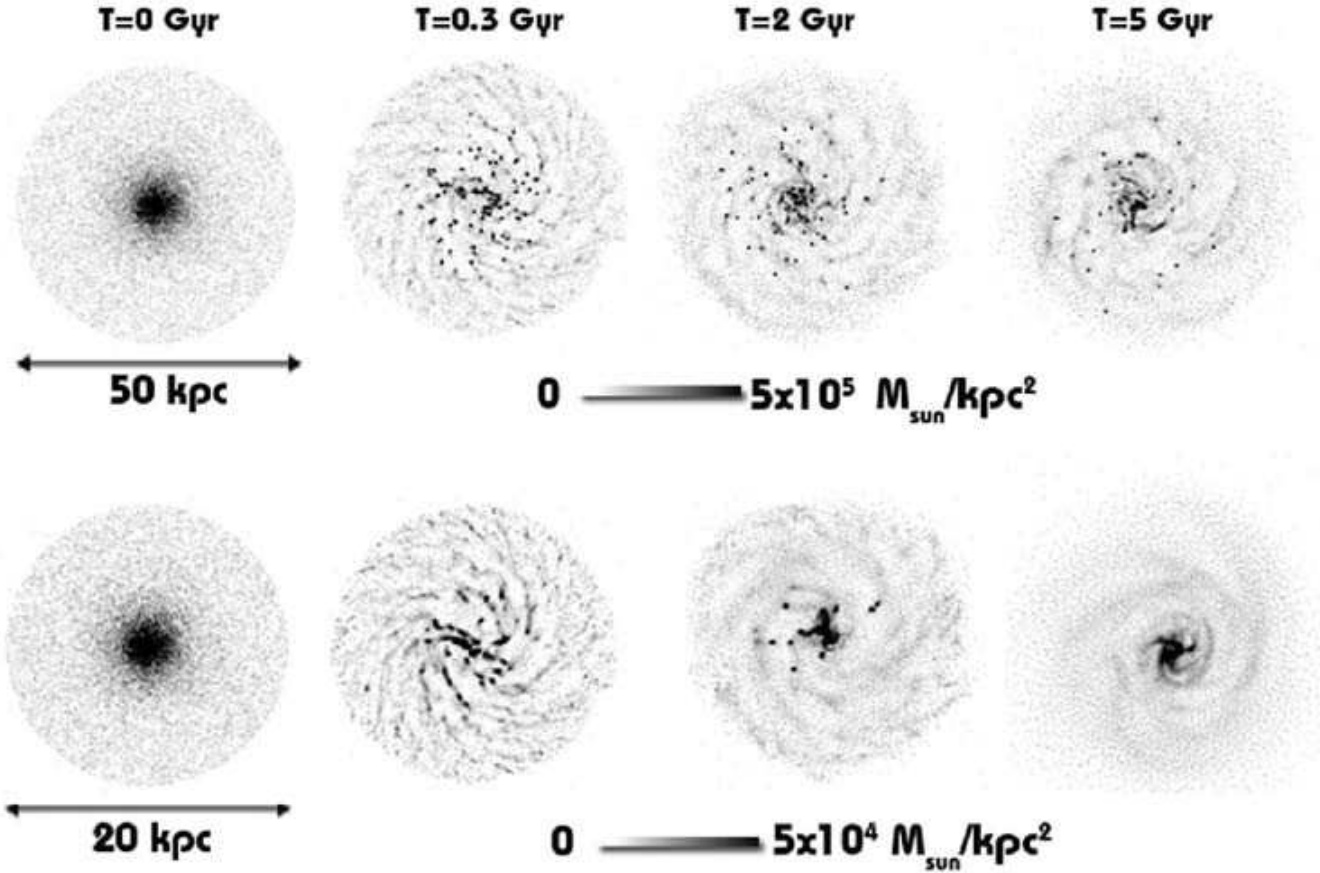}
\caption{Evolution of the gas particles in the disk for model
galaxy AM (upper panel) and BM (lower panel). The star forming
overdensities at T=0.3 Gyr and the number decrease of this regions
at T=2 (2 Gyr) and T=5 (5 Gyr) can be seen.} \label{EvolutionAM}
\end{figure*}

The goal of this work is to investigate the overall SFR evolution
of interacting galaxies by varying the spatial alignment and the
impact parameter of the interacting galaxies. Therefore firstly
the SFR of the isolated galaxies has to be studied, to distinguish
which contribution to the global SFR comes from the undisturbed
galaxies and which from the interaction. Another major point in
numerical simulations is the resolution of the system, ie. the
number of particles. Local gas densities define the amount of
newly formed stars, therefore resolution and gravitational
softening influences the quantitative maximum of the SFR. As we
are interested in the relative change of the integrated overall
SFR for interacting systems, the absolute value of the maximum of
the SFR does not influence the results, as long as the resolution,
softenings and feedback parameters are constant for all
simulations. Figure \ref{SFRrates for the isolated model galaxies}
gives the evolution of the SFR for the isolated galaxy A with
different resolutions (see Table
\ref{galaxyproperties_resolution}). The maximum of the SFR at t
$\sim$ 0.3 Gyr is caused to instabilities on the gaseous disk as
it begins to rotate and evolves. A detailed stability analysis for
disks with different gas masses and equation of state softenings
can be found in Springel et al. (2004, section 6). Integrating and
normalizing the SFR over a time range of 5 Gyr results for model
Galaxy AH 1, for model Galaxy AM 0.7929 and for model Galaxy AL
0.9291. The SFR of a model galaxy with $v_{200}=80$ km/s is two
orders of magnitude lower than a model galaxy with $v_{200}=160$.
As $v_{200}$ defines the total mass and the size of a system, see
equation 1 and 2, the gas content in the disks of model galaxy A
and B differ in the order of one magnitude, see Table
\ref{galaxyproperties}. As a consequence of the lower gas mass and
the smaller disk of model galaxy B the induced instabilities
result in a smaller overall SFR. Figure \ref{EvolutionAM} shows
the gaseous and stellar matter of the isolated model galaxy AM at
t=0.5, t=2 and t=4 Gyr.

\begin{figure}[h]
\centering
\includegraphics[width=8cm]{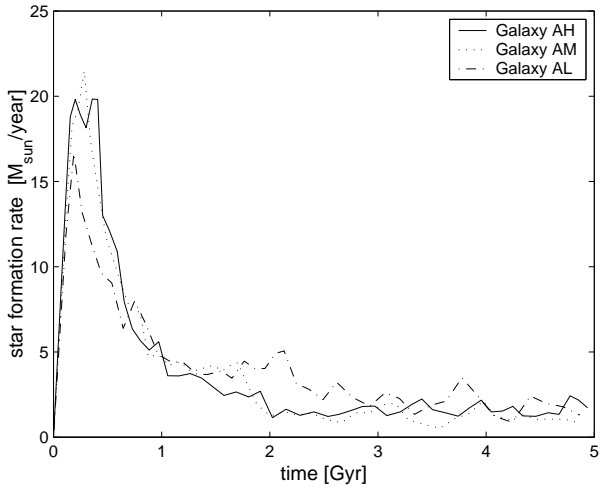}
\caption{SFRs for the isolated model galaxies. The labelling
Galaxy AM, Galaxy AH and Galaxy AL corresponds to the galaxy
properties defined in Tables \ref{galaxyproperties} and
\ref{galaxyproperties_resolution}.} \label{SFRrates for the
isolated model galaxies}
\end{figure}

\begin{figure}[h]
\includegraphics[width=8cm]{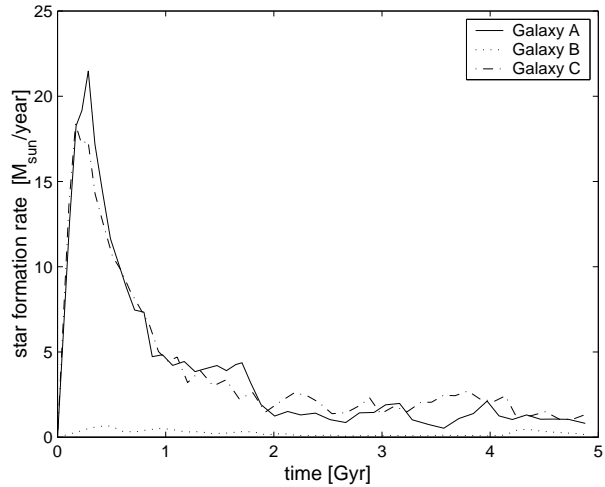}
\caption{SFRs for all isolated model galaxies given in Table
\ref{galaxyproperties} (resolution M, see Table
\ref{galaxyproperties_resolution}).} \label{SFR-isolated-galaxies}
\end{figure}

\begin{figure*}
\begin{center}
\includegraphics[width=\textwidth]{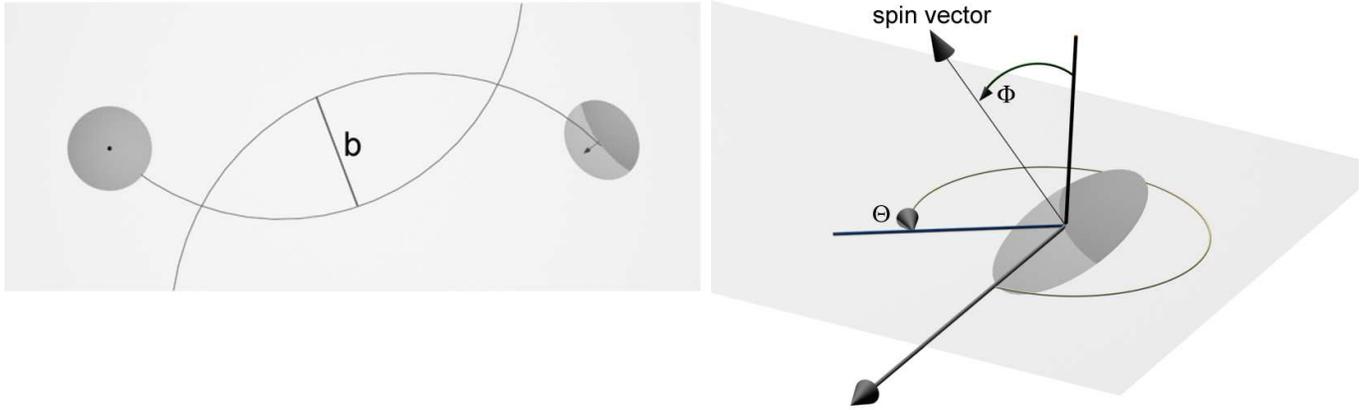}
\caption{Interaction geometry. The galaxies' positions and
velocities are chosen in such a way that the galaxies are point
masses moving on Keplerian orbits. See Duc et al. (2000) for
further descriptions.} \label{trajectory}
\end{center}
\end{figure*}

\section{The spatial alignments and impact parameters}

The geometry of the interaction was set up in the same coordinate
system as in Duc et al. (2000). Figure \ref{trajectory} shows the
angles and the trajectories. The galaxies' positions and
velocities were set up as if they were point masses on Keplerian
orbits with a minimum separation $\bf{b}$. Tables \ref{big_table1}
and \ref{big_table2} list the different angles and impact
parameters for our simulations. The alignments were chosen to
cover as much interaction geometries as possible. Of course
computational time is the limiting factor, therefore the number of
different alignments is limited. Special emphasis was put on the
investigation of co- and counter-rotating cases and on increasing
the minimum separation $\bf{b}$ for simulating fly-bys. All
together we did 9 different spatial alignments and 8 different
minimum separations for 4 different interacting systems. This
leads to an overall of 56 simulations to cover as many interacting
scenarios as possible.

\begin{landscape}

\begin{table}
\begin{center}
\caption[]{Interaction parameters for simulations 1-28 and
results. Col.2: model galaxy 1; Col. 5: model galaxy 2; Col. 8:
maximum separation [kpc]; Col. 9: minimum separation [kpc]; Col.
10: normalized total SF; Col. 11: total SF $<$ 2 Gyr; Col. 12:
total SF between 2 Gyr and 3 Gyr; Col. 13: total SF $\geq$ 3 Gyr;
Col. 14: gas/stellar mass @ t=0 Gyr; Col. 15: gas/stellar mass @
t=2 Gyr; Col. 16: gas/stellar mass @ t=3 Gyr; Col. 17: gas/stellar
mass @ t=5 Gyr; Col. 18: 80\% gas mass within radius [kpc]; Col.
19: 80\% stellar mass within radius [kpc]; Col. 20: 80\% new
formed stars within radius [kpc]}
\begin{tabular}{c c c c c c c c c c c c c c c c c c c c}
\hline  \hline &  & & &  & & & & & & & & & & & & & & & \cr
Simulation & C 2 & $\Phi_{1}$ & $\Theta_{1}$ & C 5 & $\Phi_{2}$ &
$\Theta_{2}$ & C 8 & C 9  & C 10 & C 11 & C 12 & C 13 & C 14 & C
15 & C 16 & C 17 & C 18 & C 19 & C 20\cr \hline two isolated & & &
& & & & & & & & & & & & & \cr galaxies A & A & - & - & A & - & - &
- & - & 1.00$^{1}$ & 0.79 & 0.08 & 0.13 & 0.25 & 0.19 & 0.19 &
0.18 & -$^{3}$ & -$^{3}$ & -$^{3}$ \cr Sim 1 & A & 0 & 0 & A & 0 &
0 & 200 & 0 & 2.79 & 0.61 & 0.14 & 0.25 & 0.25 & 0.14 & 0.12 &
0.08 & 5.42 & 7.23 & 1.20 \cr Sim 2 & A & 0 & 0 & A & 0 & 0 & 200
& 5 & 2.75 & 0.45 & 0.07 & 0.48 & 0.25 & 0.16 & 0.15 & 0.07 & 1.06
& 7.45 & 2.13 \cr Sim 3 & A & 0 & 0 & A & 0 & 0 & 200 & 10 & 2.60
& 0.53 & 0.06 & 0.41 & 0.25 & 0.16 & 0.15 & 0.07 & 2.45 & 4.23 &
1.25 \cr Sim 4 & A & 0 & 0 & A & 0 & 0 & 200 & 20 & 2.75 & 0.50 &
0.01 & 0.49 & 0.25 & 0.17 & 0.16 & 0.07 & 1.64 & 2.82 & 1.63 \cr
Sim 5 & A & 0 & 0 & A & 0 & 0 & 200 & 25 & 2.65 & 0.41 & 0.06 &
0.53 & 0.25 & 0.18 & 0.17 & 0.08 & 0.85 & 7.97 & 1.42 \cr Sim 6 &
A & 0 & 0 & A & 0 & 0 & 200 & 30 & 1.85 & 0.54 & 0.07 & 0.39 &
0.25 & 0.18 & 0.17 & 0.13 & 0.08 & 1.23 & 0.28 \cr Sim 7 & A & 0 &
0 & A & 0 & 0 & 200 & 40 & 1.24 & 0.78 & 0.11 & 0.11 & 0.25 & 0.19
& 0.18 & 0.17 & -$^{4}$ & -$^{4}$ & -$^{4}$ \cr Sim 8 & A & 0 & 0
& A & 0 & 0 & 200 & 50 & 1.09 & 0.72 & 0.11 & 0.16 & 0.25 & 0.19 &
0.18 & 0.17 & -$^{4}$ & -$^{4}$ & -$^{4}$ \cr Sim 9 & A & 0 & 0 &
A & 90 & 0 & 200 & 5 & 1.77 & 0.57 & 0.06 & 0.35 & 0.25 & 0.18 &
0.17 & 0.13 & 4.11 & 6.17 & 3.08 \cr Sim 10 & A & 0 & 0 & A & 90 &
90 & 200 & 5 & 2.45 & 0.47 & 0.04 & 0.49 & 0.25 & 0.18 & 0.17 &
0.10 & 1.78 & 7.12 & 1.78 \cr Sim 11 & A & 90 & 0 & A & 90 & 0 &
200 & 5 & 1.13 & 0.81 & 0.09 & 0.10 & 0.25 & 0.19 & 0.18 & 0.17 &
2.22 & 4.44 & 2.22 \cr Sim 12 & A & 90 & 0 & A & -90 & 0 & 200 & 5
& 2.09 & 0.45 & 0.07 & 0.48 & 0.25 & 0.19 & 0.18 & 0.12 & 0.10 &
6.02 & 0.54 \cr Sim 13 & A & 0 & 0 & A & 45 & 0 & 200 & 5 & 1.77 &
0.58 & 0.07 & 0.35 & 0.25 & 0.18 & 0.17 & 0.13 & 5.03 & 6.03 &
3.01 \cr Sim 14 & A & 45 & 0 & A & 45 & 0 & 200 & 5 & 1.44 & 0.65
& 0.07 & 0.28 & 0.25 & 0.18 & 0.17 & 0.15 & 1.69 & 5.09 & 0.85 \cr
Sim 15 & A & -45 & 0 & A & 45 & 0 & 200 & 5 & 2.12 & 0.45 & 0.05 &
0.50 & 0.25 & 0.18 & 0.17 & 0.11 & 0.70 & 6.34 & 1.17 \cr Sim 16 &
A & 0 & 0 & A & 180 & 0 & 200 & 5 & 2.55 & 0.55 & 0.00 & 0.45 &
0.25 & 0.17 & 0.16 & 0.09 & 0.13 & 8.03 & 0.76 \cr \hline isolated
galaxies & & & & & & & & & & & & & & & & \cr A and B & A & - & - &
B & - & - & - & - & 1.00$^{2}$ & 0.79 & 0.08 & 0.13 & 0.25 & 0.19
& 0.19 & 0.18 & -$^{3}$ & -$^{3}$ & -$^{3}$ \cr Sim 17 & A & 0 & 0
& B & 0 & 0 & 200 & 0 & 1.33 & 0.77 & 0.12 & 0.11 & 0.25 & 0.19 &
0.18 & 0.17 & 14.06 & 8.88 & 4.07 \cr Sim 18 & A & 0 & 0 & B & 0 &
0 & 200 & 5 & 1.19 & 0.78 & 0.07 & 0.15 & 0.25 & 0.20 & 0.20 &
0.19 & 20.05 & 11.34 & 7.91 \cr Sim 19 & A & 0 & 0 & B & 0 & 0 &
200 & 25 & 1.13 & 0.73 & 0.11 & 0.16 & 0.25 & 0.20 & 0.20 & 0.19 &
-$^{4}$ & -$^{4}$ & -$^{4}$\cr Sim 20 & A & 0 & 0 & B & 0 & 0 &
200 & 50 & 1.11 & 0.74 & 0.12 & 0.14 & 0.25 & 0.20 & 0.20 & 0.19 &
-$^{4}$ & -$^{4}$ & -$^{4}$ \cr Sim 21 & A & 0 & 0 & B & 90 & 0 &
200 & 5 & 1.11 & 0.72 & 0.10 & 0.18 & 0.25 & 0.20 & 0.20 & 0.19 &
21.15 & 12.08 & 8.63\cr Sim 22 & A & 0 & 0 & B & 90 & 90 & 200 & 5
& 1.28 & 0.68 & 0.12 & 0.20 & 0.25 & 0.20 & 0.19 & 0.18 & 16.04 &
10.13 & 4.60 \cr Sim 23 & A & 90 & 0 & B & 90 & 0 & 200 & 5 & 1.22
& 0.73 & 0.14 & 0.13 & 0.25 & 0.20 & 0.19 & 0.18 & 10.51 & 8.18 &
8.18 \cr Sim 24 & A & 90 & 0 & B & -90 & 0 & 200 & 5 & 1.18 & 0.85
& 0.00 & 0.15 & 0.25 & 0.20 & 0.19 & 0.18 & 12.85 & 9.64 & 8.84
\cr Sim 25 & A & 0 & 0 & B & 45 & 0 & 200 & 5 & 1.13 & 0.72 & 0.08
& 0.20 & 0.25 & 0.20 & 0.20 & 0.19 & 20.85 & 12.75 & 9.66 \cr Sim
26 & A & 45 & 0 & B & 45 & 0 & 200 & 5 & 1.23 & 0.71 & 0.12 & 0.17
& 0.25 & 0.20 & 0.19 & 0.18 & 15.40 & 9.86 & 8.63 \cr Sim 27 & A &
-45 & 0 & B & 45 & 0 & 200 & 5 & 1.31 & 0.70 & 0.11 & 0.19 & 0.25
& 0.20 & 0.19 & 0.18 & 16.00 & 9.43 & 6.56 \cr Sim 28 & A & 0 & 0
& B & 180 & 0 & 200 & 5 & 1.28 & 0.51 & 0.10 & 0.39 & 0.25 & 0.20
& 0.20 & 0.18 & 5.39 & 8.56 & 2.04\cr \hline

\end{tabular}
\label{big_table1}
\end{center}
$^{1}$....$2\times\int_{t\,=\,0\,Gyr}^{t\,=\,5\,Gyr}SFR_A(t)dt=4.4\times
10^{9} $M$_{\sun}$, this is normalised to 1.\\
$^{2}$....$\int_{t\,=\,0\,Gyr}^{t\,=\,5\,Gyr}SFR_A(t)dt+
\int_{t\,=\,0\,Gyr}^{t\,=\,5\,Gyr}SFR_B(t)dt=2.4\times
10^{9}$M$_{\sun}$, this is normalised to 1.\\
$^{3}$....see Table \ref{Cutoff radii} for numbers.\\
$^{4}$....no merger occurs within simulation time, therefore no
cut-off radii are given.\\
\end{table}

\end{landscape}

\begin{landscape}

\begin{table}
\begin{center}
\caption[]{Interaction parameters for simulations 29-56 and
results. Col.2: model galaxy 1; Col. 5: model galaxy 2; Col. 8:
maximum separation [kpc]; Col. 9: minimum separation [kpc]; Col.
10: normalized total SF; Col. 11: total SF $<$ 2 Gyr; Col. 12:
total SF between 2 Gyr and 3 Gyr; Col. 13: total SF $\geq$ 3 Gyr;
Col. 14: gas/stellar mass @ t=0 Gyr; Col. 15: gas/stellar mass @
t=2 Gyr; Col. 16: gas/stellar mass @ t=3 Gyr; Col. 17: gas/stellar
mass @ t=5 Gyr; Col. 18: 80\% gas mass within radius [kpc]; Col.
19: 80\% stellar mass within radius [kpc]; Col. 20: 80\% new
formed stars within radius [kpc]}
\begin{tabular}{c c c c c c c c c c c c c c c c c c c c}
\hline  \hline &  & & &  & & & & & & & & & & & & & & & \cr
Simulation & C 2 & $\Phi_{1}$ & $\Theta_{1}$ & C 5 & $\Phi_{2}$ &
$\Theta_{2}$ & C 8 & C 9  & C 10 & C 11 & C 12 & C 13 & C 14 & C
15 & C 16 & C 17 & C 18 & C 19 & C 20\cr \hline two isolated & & &
& & & & & & & & & & & & & \cr galaxies A and C & A & - & - & C & -
& - & - & - & 1.00$^{1}$ & 0.80 & 0.07 & 0.13 & 0.25 & 0.19 & 0.19
& 0.18 & -$^{3}$ & -$^{3}$ & -$^{3}$\cr Sim 29 & A & 0 & 0 & C & 0
& 0 & 200 & 0 & 2.77 & 0.52 & 0.10 & 0.38 & 0.25 & 0.15 & 0.14 &
0.08 & 6.91 & 8.83 & 1.92 \cr Sim 30 & A & 0 & 0 & C & 0 & 0 & 200
& 5 & 2.96 & 0.44 & 0.08 & 0.48 & 0.25 & 0.16 & 0.15 & 0.08 & 6.98
& 9.50 & 3.17 \cr Sim 31 & A & 0 & 0 & C & 0 & 0 & 200 & 10 & 2.72
& 0.44 & 0.07 & 0.49 & 0.25 & 0.17 & 0.16 & 0.08 & 5.48 & 5.48 &
5.48 \cr Sim 32 & A & 0 & 0 & C & 0 & 0 & 200 & 20 & 2.18 & 0.56 &
0.01 & 0.43 & 0.25 & 0.18 & 0.17 & 0.08 & 11.60 & 11.60 & 11.60
\cr Sim 33 & A & 0 & 0 & C & 0 & 0 & 200 & 25 & 2.69 & 0.40 & 0.06
& 0.54 & 0.25 & 0.18 & 0.17 & 0.09 & 1.36 & 8.96 & 2.71 \cr Sim 34
& A & 0 & 0 & C & 0 & 0 & 200 & 30 & 1.56 & 0.63 & 0.10 & 0.27 &
0.25 & 0.18 & 0.17 & 0.15 & 7.36 & 4.93 & 4.93 \cr Sim 35 & A & 0
& 0 & C & 0 & 0 & 200 & 40 & 1.10 & 0.77 & 0.10 & 0.13 & 0.25 &
0.19 & 0.18 & 0.17 & -$^{4}$ & -$^{4}$ & -$^{4}$ \cr Sim 36 & A &
0 & 0 & C & 0 & 0 & 200 & 50 & 1.12 & 0.86 & 0.00 & 0.14 & 0.25 &
0.19 & 0.18 & 0.17 & -$^{4}$ & -$^{4}$ & -$^{4}$ \cr Sim 37 & A &
0 & 0 & C & 90 & 0 & 200 & 5 & 1.60 & 0.70 & 0.08 & 0.22 & 0.25 &
0.17 & 0.17 & 0.14 & 10.84 & 9.64 & 6.62 \cr Sim 38 & A & 0 & 0 &
C & 90 & 90 & 200 & 5 & 2.03 & 0.51 & 0.05 & 0.44 & 0.25 & 0.18 &
0.17 & 0.11 & 5.60 & 9.56 & 4.61 \cr Sim 39 & A & 90 & 0 &C & 90 &
0 & 200 & 5 & 1.37 & 0.70 & 0.07 & 0.23 & 0.25 & 0.19 & 0.18 &
0.16 & 4.79 & 5.98 & 1.20 \cr Sim 40 & A & 90 & 0 & C & -90 & 0 &
200 & 5 & 2.08 & 0.45 & 0.05 & 0.50 & 0.25 & 0.19 & 0.18 & 0.11 &
0.53 & 6.61 & 1.06 \cr Sim 41 & A & 0 & 0 & C & 45 & 0 & 200 & 5 &
1.64 & 0.65 & 0.06 & 0.29 & 0.25 & 0.17 & 0.17 & 0.14 & 9.58 &
10.11 & 6.39 \cr Sim 42 & A & 45 & 0 & C & 45 & 0 & 200 & 5 & 1.52
& 0.66 & 0.07 & 0.27 & 0.25 & 0.18 & 0.18 & 0.16 & 1.46 & 5.84 &
2.19 \cr Sim 43 & A & -45 & 0 & C & 45 & 0 & 200 & 5 & 2.10 & 0.46
& 0.08 & 0.46 & 0.25 & 0.19 & 0.18 & 0.12 & 2.70 & 7.20 & 2.70 \cr
Sim 44 & A & 0 & 0 & C & 180 & 0 & 200 & 5 & 2.20 & 0.61 & 0.00 &
0.39 & 0.25 & 0.16 & 0.15 & 0.10 & 3.40 & 9.81 & 3.77 \cr \hline
two isolated & & & & & & & & & & & & & & & & \cr model galaxies B
& B & - & - & B & - & - & - & - & 1.00$^{2}$ & 0.62 & 0.09 & 0.30
& 0.25 & 0.22 & 0.22 & 0.21 & -$^{3}$ & -$^{3}$ & -$^{3}$ \cr Sim
45 & B & 0 & 0 & B & 0 & 0 & 200 & 0 & 4.76 & 0.43 & 0.16 & 0.41 &
0.25 & 0.18 & 0.15 & 0.09 & 0.40 & 1.59 & 0.80 \cr Sim 46 & B & 0
& 0 & B & 0 & 0 & 200 & 5 & 4.15 & 0.24 & 0.04 & 0.71 & 0.25 &
0.21 & 0.21 & 0.11 & 0.80 & 3.21 & 1.12 \cr Sim 47 & B & 0 & 0 & B
& 0 & 0 & 200 & 25 & 1.23 & 0.53 & 0.27 & 0.20 & 0.25 & 0.23 &
0.22 & 0.21 & 3.77 & 4.67 & 3.01 \cr Sim 48 & B & 0 & 0 & B & 0 &
0 & 200 & 50 & 0.96 & 0.72 & 0.00 & 0.28 & 0.25 & 0.23 & 0.23 &
0.22 & -$^{4}$ & -$^{4}$ & -$^{4}$ \cr Sim 49 & B & 0 & 0 & B & 90
& 0 & 200 & 5 & 1.92 & 0.56 & 0.12 & 0.32 & 0.25 & 0.22 & 0.21 &
0.17 & 1.25 & 2.50 & 1.07 \cr Sim 50 & B & 0 & 0 & B & 90 & 90 &
200 & 5 & 2.26 & 0.45 & 0.08 & 0.47 & 0.25 & 0.21 & 0.20 & 0.17 &
1.34 & 2.56 & 1.34 \cr Sim 51 & B & 90 & 0 & B & 90 & 0 & 200 & 5
& 1.76 & 0.56 & 0.17 & 0.27 & 0.25 & 0.22 & 0.21 & 0.19 & 1.21 &
2.11 & 0.60 \cr Sim 52 & B & 90 & 0 & B & -90 & 0 & 200 & 5 & 3.10
& 0.25 & 0.07 & 0.68 & 0.25 & 0.22 & 0.21 & 0.15 & 0.25 & 2.23 &
0.49 \cr Sim 53 & B & 0 & 0 & B & 45 & 0 & 200 & 5 & 3.35 & 0.31 &
0.05 & 0.64 & 0.25 & 0.21 & 0.20 & 0.14 & 1.23 & 2.63 & 1.40 \cr
Sim 54 & B & 45 & 0 & B & 45 & 0 & 200 & 5 & 3.29 & 0.28 & 0.06 &
0.65 & 0.25 & 0.22 & 0.21 & 0.15 & 1.28 & 2.56 & 0.64 \cr Sim 55 &
B & -45 & 0 & B & 45 & 0 & 200 & 5 & 4.14 & 0.20 & 0.06 & 0.74 &
0.25 & 0.22 & 0.21 & 0.11 & 0.09 & 2.62 & 0.60 \cr Sim 56 & B & 0
& 0 & B & 180 & 0 & 200 & 5 & 1.28 & 0.90 & 0.00 & 0.10 & 0.25 &
0.20 & 0.19 & 0.18 & -$^{4}$ & -$^{4}$ & -$^{4}$ \cr \hline
\end{tabular}
\label{big_table2}
\end{center}
$^{1}$....$2\times\int_{t\,=\,0\,Gyr}^{t\,=\,5\,Gyr}SFR_A(t)dt=4.5\times
10^{9} $M$_{\sun}$, this is normalised to 1.\\
$^{2}$....$\int_{t\,=\,0\,Gyr}^{t\,=\,5\,Gyr}SFR_A(t)dt+
\int_{t\,=\,0\,Gyr}^{t\,=\,5\,Gyr}SFR_B(t)dt=2.4\times
10^{9}$M$_{\sun}$, this is normalised to 1.\\
$^{3}$....see Table \ref{Cutoff radii} for numbers.\\
$^{4}$....no merger occurs within simulation time, therefore no
cut-off radii are given.\\
\end{table}

\end{landscape}

\section{Results of the simulations}
\subsection{Collisions between model galaxies A}

\begin{figure*}[ht]
\begin{center}
\includegraphics[width=16cm]{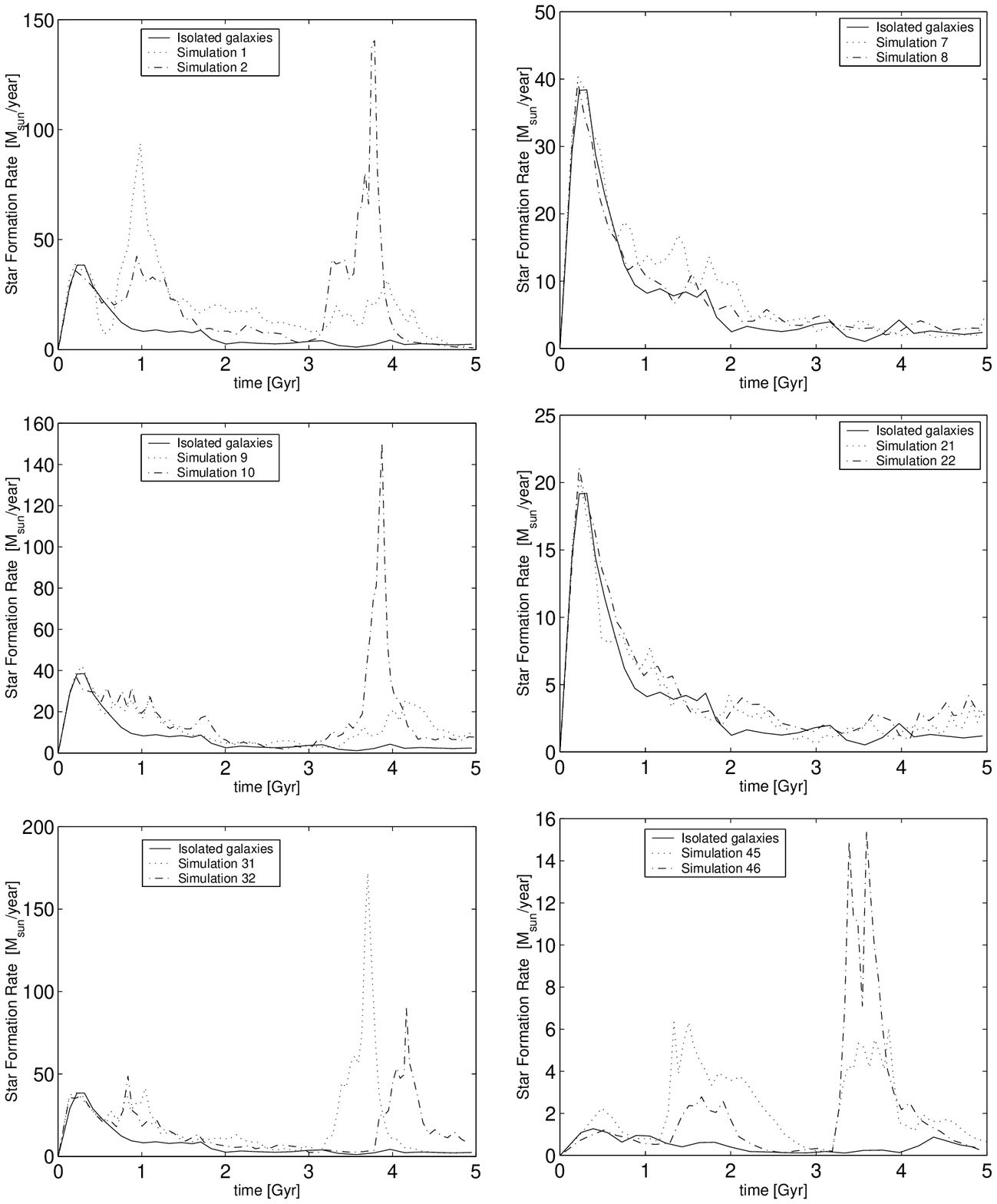}
\caption{Evolution of the SFRs for several simulations (see Tables
\ref{big_table1} and \ref{big_table2}). Simulation 1 and 2 show
the evolution of the SFRs if the minimum separation is increased.
If the minimum separation is increased from 40 to 50 kpc
(simulation 7 and 8) the SFRs do not change. In simulation 9 and
10 the big difference between two spatial alignments on the
evolution of the SFRs is shown. If one of the interacting galaxies
has 10 times less mass and does not form stars by its own
(simulation 21 and 22) the overall SFR does not change in
comparison to the isolated systems. In simulation 45 and 46 the
maximum enhancement of the ISFR of all simulations occurs.}
\label{ISFR}
\end{center}
\end{figure*}

In Table \ref{big_table1} the integrated star formation rates
(ISFR) for all collisions between two model galaxies A as well as
those for two isolated model galaxies A are listed. All ISFR for
simulations 1-16 are given in units of
$2\times\int_{t\,=\,0\,Gyr}^{t\,=\,5\,Gyr}\,SFR_A(t) dt =
4.4\times 10^{9} $M$_{\sun}$, ie. the integrated star formation
rate of two isolated model galaxies A. Thereby $SFR_A(t)$ is the
star formation rate for the isolated model galaxy A, see Fig
\ref{SFRrates for the isolated
model galaxies}.\\
All galaxy interactions show an enhancement in the ISFR. The
maximum arises in simulation 1 with a 2.79 times higher ISFR in
comparison to two isolated model galaxies A. Simulation 1 is a
co-rotation edge-on collision with a minimum separation of 0 kpc,
see Table \ref{big_table1} for a list of the interaction
parameters. If the minimum separation is increased (simulations
2-8) the ISFR decreases. A fly-by with a minimum separation of 50
kpc results in a small enhancement of the ISFR of 9\%.\\
Counter-rotating interacting systems do not always have a lower or
higher ISFR in comparison to co-rotating systems. It depends on
the spatial alignment of the interacting galaxies. While
simulations 2 and 16 show a decrease of the ISFR for
counter-rotating systems, simulations 11 and 12 show an enormous
enhancement of the ISFR for the counter-rotating system in
comparison to the co-rotating ones (185\%). Columns 11,12 and 13
of Table \ref{big_table1} give the ISFR for different time
intervals, i.e. t$<2$ Gyr, 2 $\leq$ t $<$ 3 Gyr and t$\geq$ 3 Gyr,
always relative to the ISFR for the whole simulation. These
intervals were chosen such that the first interval ends shortly
after the first encounter, the second interval covers the time
between the first and the second encounter and the last interval
starts shortly before the second encounter and lasts to the end of
the simulation. At the end of the simulation the two galaxies
always form a bound system, except in simulations 7 and 8, which
do not merge within the simulation time of 5 Gyr. In Fig.
\ref{ISFR} the evolution of the SFRs for some particular
simulations are given. Simulations 1 and 2 show the dependence of
the minimum separation on the strengths of the SFRs. Whereas in
simulation 1 (minimum separation 0 kpc) the first encounter
produces an increase of the SFR, simulation 2 (minimum separation
5 kpc) shows exactly the opposite at the first encounter. While in
simulation 1 44\% of the gas was converted into stellar matter
after the first encounter (t=2 Gyr), in simulation 2 only 36\% of
the gas was converted. The second encounter then results in a very
high SFR for simulation 2, whereas simulation 1 does not show such
an enormous enhancement (see Fig. \ref{ISFR}). After the first and
before the second encounter the SFR in simulation 1 does not
decrease to values as in simulations 2-16 (see Table
\ref{big_table1}, Col. 12).\\
If the minimum separation is greater than 30 kpc the SFRs increase
only by a factor of 24\% (simulation 7) and 9\% (simulation 8).
Simulations 9 and 10 (see Fig. \ref{ISFR} show the big difference
between different spatial alignments but same minimum and maximum
separation on the evolution of the SFR. The first case where an
edge-on model galaxy A interacts with a face-on model galaxy A
results in an enhancement of the ISFR by a factor of 1.75. If we
change in addition $\Theta_{2}$ to $90^{\circ}$ the ISFR increases
by a factor of 2.45. This leads to the conclusion that not only
the minimum separation influences the ISFR, but that also the
spatial alignments are an important factor. The SFR of the
isolated model galaxies A is always given for reference in Tables
\ref{big_table1} and \ref{big_table2}.

\subsection{Collisions between model galaxies A and B}

The involved galaxies A and B have a mass ratio of 8:1, see Table
\ref{galaxyproperties}. The ISFR for the collisions between the
model galaxies A and B show that there is only an average increase
of by a factor of 1.21 in comparison to the ISFR of the two
isolated model galaxies A and B. For collisions between A and B
all ISFR are given in units of $\int_{t=0 Gyr}^{t=5 Gyr} SFR_A(t)
dt+\int_{t=0 Gyr}^{t=5 Gyr} SFR_B(t) dt \approx 2.2\times
10^{9}$M$_{\sun}$. Note that the ISFR of the isolated model galaxy
B is below 0.5\% of the ISFR of the isolated model galaxy A and
therefore negligible. Figure \ref{ISFR} shows the evolution of the
SFRs of simulations 21 and 22 (see Table \ref{big_table1} for
details). As galaxy A has about 8 times more gas than galaxy B the
collisions do not increase the ISFR like collisions between two
model galaxies A. The SFR of model galaxy A yields therefore the
major contribution to the ISFR of simulations 17-28. This leads to
the conclusion that a merger of a gas-poor galaxy with a large
galaxy like model galaxy A does not lead to a large increase of
the SFR. The major effect that we see in our simulations between
model galaxy A and B is the redistribution of gaseous and stellar
matter in huge spaces around the interacting system. In section 8
this result will be presented in detail.

\subsection{Collisions between model galaxies A and C}

Simulations 29-44 each involve one system with and one without a
bulge (model galaxy A and C). The chosen bulge properties, see
Table \ref{galaxyproperties}, do not dramatically influence the
evolution of the SFRs. Comparing the evolution of the SFRs of the
isolated galaxy A (without bulge) and isolated galaxy C (with
bulge) does not yield a significant difference, see Fig.
\ref{SFR-isolated-galaxies}. ISFR are again given in units of
$\int_{t=0 Gyr}^{t=5 Gyr} SFR_A(t) dt+\int_{t=0 Gyr}^{t=5 Gyr}
SFR_C(t) dt=4.5\times 10^{9}$M$_{\sun}$, see Fig. \ref{SFRrates
for the isolated model galaxies}. The results show approximately
the same behavior as the results of simulations 1-16 (collisions
between two model galaxies A). In Fig. \ref{ISFR} the SFRs of
simulations 31 and 32 are presented as an example. The decrease of
the ISFR by increasing the minimum separation is nearly identical
to that of simulations 1-8.

\subsection{Collisions between model galaxies B}

Collisions between two model galaxies B show the strongest
enhancement of ISFR. The maximum enhancement is 4.76 times higher
than for the isolated model galaxies. Note that model galaxy B is
chosen in such a way that the isolated galaxy show hardly any star
formation, see Fig. \ref{SFR-isolated-galaxies}. In case of close
interaction the gas can exceed locally the defined threshold for
star formation and therefore a dramatic increase of the SFR
occurs. Similar to all other simulations, simulations 45,46,47 and
48 show a decrease of the ISFR with increasing minimum separation.
Simulation 48 (minimum separation 50 kpc) even results in a
decrease of 4\% of the ISFR in comparison to the isolated model
galaxies. Two isolated model galaxies B form $2.8\times
10^{8}$M$_{\sun}$ of stars in 5 Gyr. In Fig. \ref{ISFR} the
evolution of the SFRs of simulations 45 and 46 are shown as an
example.\\
It is important to point out that interaction of small spiral
galaxies can produce strong star formation in comparison to the
isolated systems and therefore make them easily observable. They
might be good tracers for the frequency of galaxy-galaxy
interaction in the distant universe.

\section{The gas to star transformation efficiency}

Different simulations show different star formation rates and
ISFR, therefore they are more or less efficient in transforming
gas into stellar matter. The ISFR can be dominated by an extreme
starburst-like event for a short duration (several million years)
or a long term low enhancement of SFR due to interaction. As
observers measure quantities like gas mass or stellar
mass/luminosities of a galaxy in a certain evolutionary state of a
galaxy, the knowledge of the gas/stellar ratios provides a crucial
link between observation and theory. Therefore we elaborate in
this section purely on the gas/stellar mass ratios of our
simulations.

\subsection{Collisions between model galaxies A}

\begin{figure}[ht]
\begin{center}
\includegraphics[width=8.8cm]{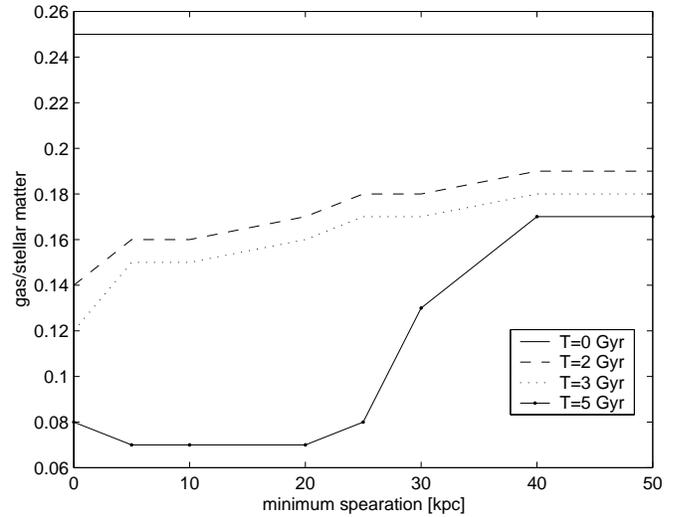}
\caption{Gas/stellar mass ratio for simulations 1-8 at simulation
times 0,2,3 and 5 Gyr} \label{gas/stellar AA}
\end{center}
\end{figure}

For all simulations the ratio of gaseous matter and stellar matter
at certain timesteps were calculated. In Table \ref{big_table1}
the gas/stellar mass ratios for simulations 1-16 are given. The
ratios are always given at the beginning of each simulation (t=0
Gyr), at t=2 Gyr, t=3 Gyr and at t=5 Gyr (end of all simulations).
The timesteps were chosen in such a way that they correspond to
shortly after the first encounter, shortly before the second
encounter and shortly after the second encounter. Except for
simulations 7 and 8 all collisions between two model galaxies A
end up in a single elliptical galaxy, i.e. they merge. Simulations
7 and 8 do not merge within the simulation time of 5 Gyr, they are
just fly-bys with only one encounter. The highest efficiency in
converting gaseous matter into stars are found in simulations 1-5.
Simulation 16, a counter-rotating collision, shows less efficiency
in gaseous to stellar matter conversion than simulation 2, the
co-rotating encounter with the identical interaction geometry.
Figure \ref{gas/stellar AA} shows the ratios of gaseous to stellar
matter for simulations 1-8 (increasing minimum separation) for
different times. At the beginning of the simulations, 25\% of all
galaxies' total disk matter is gas. The efficiency of gaseous to
stellar mass conversion follows a nearly linear behavior for
simulations 1-8 after the first encounter. The best linear fit for
the efficiency after the first encounter is

\begin{equation}
e(r)=0.94\times10^{-3}\,r+0.15
\end{equation}
where $e$ is the ratio of gaseous to stellar matter and r is the
minimum separation in kpc. The small slope of $0.94\times10^{-3}
(\textrm{kpc})^{-1}$ shows that there is no strong dependence of
$e$ on the minimum separation after the first encounter in our
model. This changes dramatically after the second encounter (t=5
Gyr). For simulations 1-4 there is the same, nearly constant,
behaviour of $e$. This results from the fact, that for a minimum
separation above 40 kpc no merger occurs within the simulation
time. For simulations 1-8 we conclude:

\begin{itemize}
\item{a minimum separation below 20 kpc ($\sim 6\times$R$_{d}$)
results in an efficiency, which scales nearly linear with r (the
minimum separation). This holds for the first and the second
encounter. The slope is very flat, therefore no strong dependence
is found.} \item{for minimum separations above 40 kpc there is
only one encounter over the whole simulation time of 5 Gyr.
Therefore no enhanced star formation due to galaxy-galaxy
interaction occurs.} \item{if the minimum separation is below
$\sim 6\times$R$_{d}$ the quotient $e$ is almost the same for the
first and the second encounter, i.e. $\sim$40\% of the gas is
converted into stars at each collision.}
\end{itemize}

\noindent Figure \ref{gas/stellar AAPHI2} shows the gas to stellar
ratios for simulations 2,9,13 and 16 (increasing $\Phi$ for one
member of the system). If $\Phi_2$ is increased to values of about
90$^{\circ}$ we find a decrease of the ISFR. If we go beyond
$90^{\circ}$ the ISFR increases and therefore the ratios of
gaseous to stellar matter decreases. Simulations 2 and 16 show
different ratios, this leads to the conclusion that a co-rotating
encounter has a higher ISFR than a counter-rotating one. This can
be seen in almost all co-rotating counter-rotating simulations,
see Table \ref{big_table1}. Only simulations 11 and 12 do not
comply with this pattern. The results show in addition a greater
dependence of the conversion of gaseous into stellar matter on
$\Phi_{2}$ after the second encounter (see Fig. \ref{gas/stellar
AAPHI2}). Simulations with $45^{\circ}<\Phi_2<90^{\circ}$ do have
a lower ISFR after the first encounter, which increases by a
factor of 2 after the second encounter.

\begin{figure}[]
\begin{center}
\includegraphics[width=8.8cm]{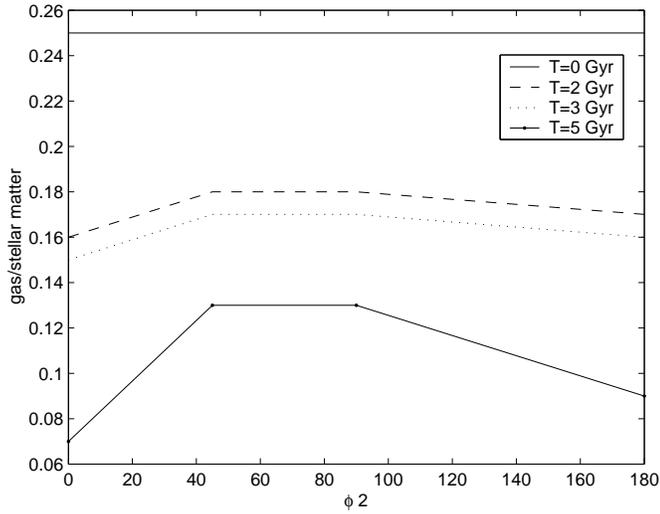}
\caption{Gas/stellar mass ratio for simulations with increasing
$\Phi_2$ at simulation times 0,2,3 and 5 Gyr} \label{gas/stellar
AAPHI2}
\end{center}
\end{figure}

\subsection{Collisions between model galaxies A and B}

Simulations 17-28, interaction between model galaxy A and B, do
not show a significant difference to the isolated systems. This is
due to the fact that the less massive partner in the interacting
system has not enough gaseous and stellar matter to disturb the
normal evolution of the more massive partner (model galaxy A).
However, the smaller partner of the interacting system
redistributes the gas and stellar components of both galaxies up
to very large distances (several hundred kpc) from the centre of
baryonic mass. The next section will go into more details on that.
Table \ref{big_table1} lists all the numbers.

\subsection{Collisions between model galaxies A and C}

The introduction of a spiral galaxy with a bulge (model galaxy C)
does not change the efficiency of converting gaseous into stellar
matter dramatically. Only at very small minimal separations ($<5$
kpc) the bulge seems to decrease the efficiency, see Table
\ref{big_table2}. The best linear fit for the efficiency after the
first encounter is

\begin{equation}
e(r)=0.76\times10^{-3}\,r+0.16
\end{equation}
where $e$ is the ratio of gaseous and stellar matter and r is the
minimum separation in kpc. The difference to equation (9) is very
small. After the second encounter the efficiency for transferring
gaseous into stellar matter shows the same dependencies on the
minimum separation as in simulations
1-8.\\
Changes of $\Phi_2$ show the same  behavior as collisions between
two model galaxies A. Again the decrease of the efficiency to
convert gaseous matter into stellar matter for collisions with
$45^{\circ}<\Phi_2<90^{\circ}$ can be seen in Fig.
\ref{gas/stellar ACPHI2}.

\begin{figure}[]
\begin{center}
\includegraphics[width=8.8cm]{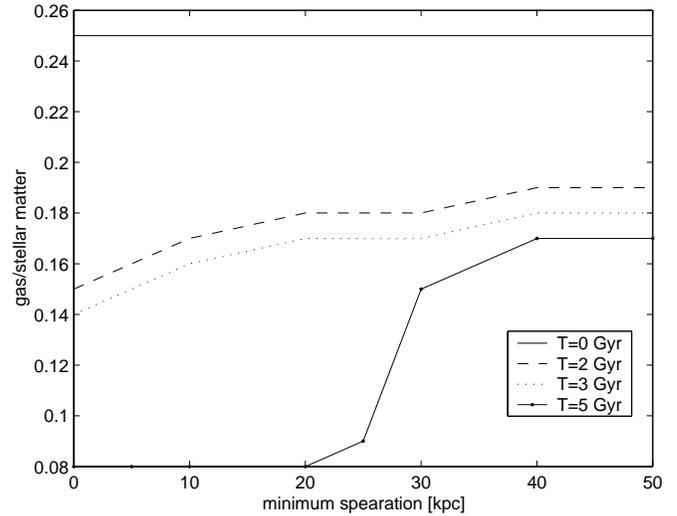}
\caption{Gas/stellar mass ratio for simulations 29-36 at
simulation times 0,2,3 and 5 Gyr} \label{gas/stellar AC}
\end{center}
\end{figure}

\begin{figure}[]
\begin{center}
\includegraphics[width=8.8cm]{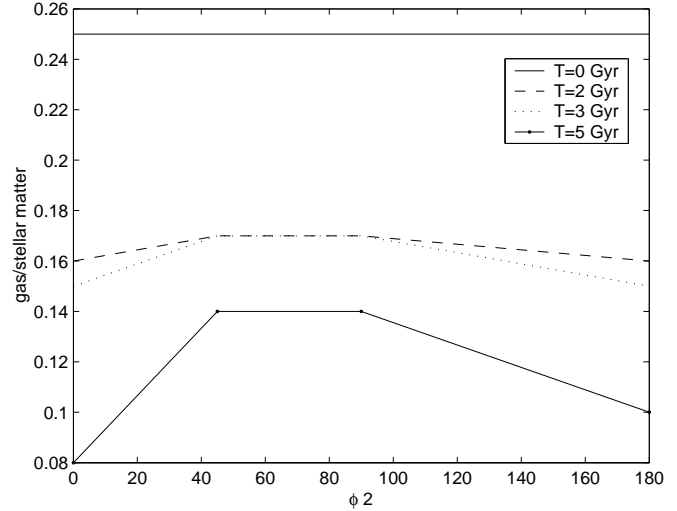}
\caption{Gas/stellar mass ratio for simulations with increasing
$\Phi_2$ at simulation time 0,2,3 and 5 Gyr} \label{gas/stellar
ACPHI2}
\end{center}
\end{figure}

\subsection{Collisions between model galaxies B}

\begin{figure}[]
\begin{center}
\includegraphics[width=8cm]{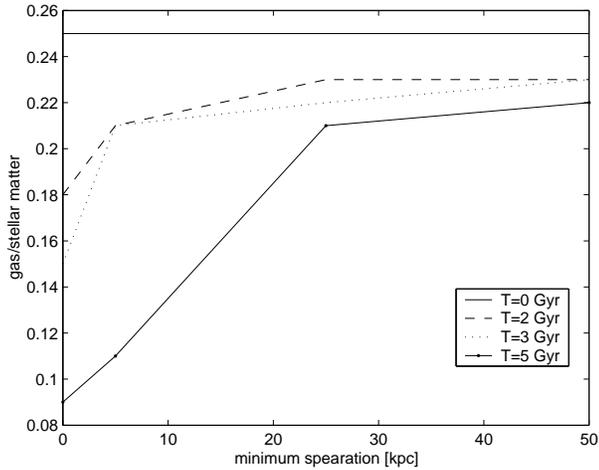}
\caption{Gas/stellar mass ratio for simulations 45-48 at
simulation times 0,2,3 and 5 Gyr} \label{gas/stellar BB}
\end{center}
\end{figure}

As a consequence of the low mass ($\sim\frac{1}{8}$ of model
galaxy A and C) the disk scale length R$_{d}$ for the undisturbed
galaxy B is 2.25 $h^{-1}$ kpc. For this reason the efficiency for
converting gaseous into stellar matter decreases with increasing
minimum separation r very fast (see Fig. \ref{gas/stellar BB}).
For r $>$ 25 kpc almost no enhancement of the ISFR is observable.
A close encounter $0 < $r$ < $5 kpc gives almost the same relative
efficiency in converting gas to stellar matter as collisions
between two model galaxies A or between galaxies A and C do.

In contrast to that changes of $\Phi_{2}$ result in different
efficiencies than the collisions between two model galaxies A
(respectively galaxies A and C), see Fig. \ref{gas/stellar
BBPHI2}. After the first encounter (t=2 Gyr) no significant change
of the efficiency can be seen, but at the end of the simulation
time (t=5 Gyr) the efficiency decreases rapidly with increasing
$\Phi_{2}$. It seems that a merger of counterrotating model
galaxies B does not cause an enhancement of the star formation
after the second encounter. The best efficiency is given in the
edge on collisions, $\Phi_{2}$=0, with a 2 times higher efficiency
after the first encounter than after the second one.

\begin{figure}[]
\begin{center}
\includegraphics[width=8cm]{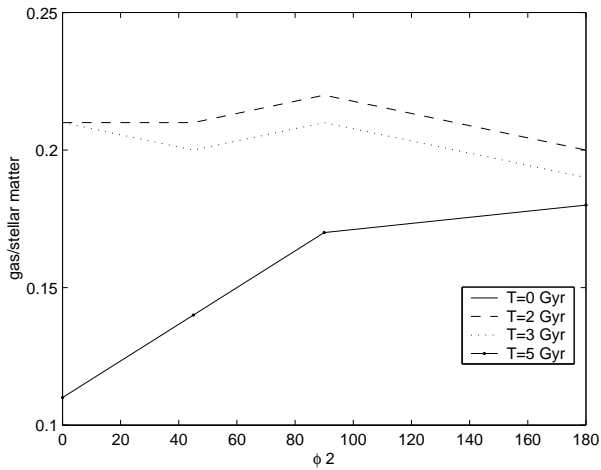}
\caption{Gas/stellar mass ratio for simulations with increasing
$\Phi_2$ at simulation times 0,2,3 and 5 Gyr} \label{gas/stellar
BBPHI2}
\end{center}
\end{figure}

\section{Spatial distribution of gas and stars}

\begin{figure}[t]
\begin{center}
\includegraphics[width=8cm]{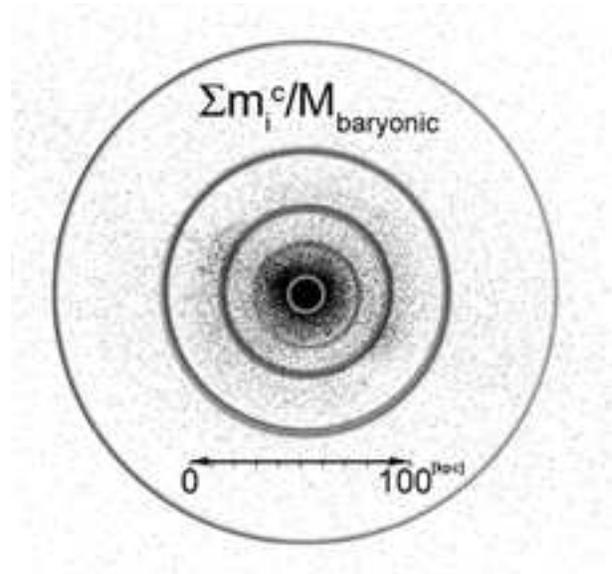}
\caption{Determination of mass profiles. The rings are centred
around the centre of mass of baryonic matter. In each ring the sum
over each particle type (stellar, gaseous and newly formed stars)
is divided by the total baryonic mass.} \label{rings}
\end{center}
\end{figure}

New observations and N-body/hydrodynamic simulations do give
evidence for an intra-cluster stellar population (ICSP) (Arnaboldi
et al. 2003; Murante et al. 2004). Arnaboldi et al. (2003) claim
that 10\% - 40\% of all stars in a galaxy cluster are members of
the ICSP. As stars are the hatchery of metals, the ICSP enriches
the intra-cluster medium (ICM) directly with metals
and energy, see Domainko et al. (2004) and references therein.\\
It is well known that galaxy-galaxy interactions can enrich the
ICM due to strong galactic winds (De Young 1978). Our simulations
show beside that another direct enrichment mechanism: vast spatial
gaseous and stellar matter distributions as a consequence of
galaxy collisions. In Tables \ref{big_table1} and \ref{big_table2}
we list the radii within which 80\% of the gaseous and stellar
matter reside after t=5 Gyr for all carried out simulations. The
cut off radii are determined in such way, that in annuli with
increasing radii around the centre of mass each particle type
(stellar, gaseous and newly formed stars) was integrated. If the
sum of a component exceeds 80\% of the overall sum of the
component, the radius was taken as the cutoff radius (see Fig.
\ref{rings}).

\noindent The total baryonic masses for the simulations are listed
in Table \ref{total baryonic masses}.

\begin{table}
\begin{center}
\caption[]{Total baryonic masses for all simulations}
\begin{tabular}{c c}
\hline \hline & Total baryonic mass \cr Simulation & [$1\times
10^{10}h^{-1}M_{\sun}$] \cr\hline 1-16 & 9.5240 \cr 17-28 & 5.3573
\cr 29-44 & 11.9050 \cr 45-56 & 1.1905 \cr \hline
\end{tabular}
\label{total baryonic masses}
\end{center}
\end{table}

\begin{figure}[t]
\begin{center}
\includegraphics[width=8cm]{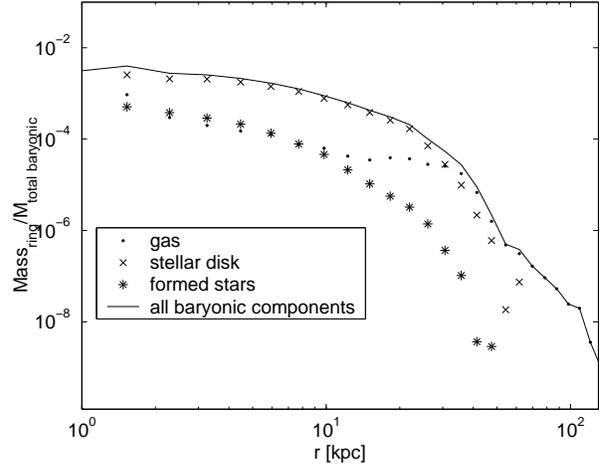}
\caption{Mass profiles of the baryonic components of the isolated
model galaxy A as a function of radius. The masses are given in
ratios of each component to the total baryonic mass of the whole
system. The binning geometry is given in Fig. \ref{rings}. The
centre of the binning is the centre of mass of the baryonic
component. The profiles are given for the last timestep in our
simulation (t=5 Gyr).} \label{profile_AM}
\end{center}
\end{figure}

\begin{figure}[t]
\begin{center}
\includegraphics[width=8cm]{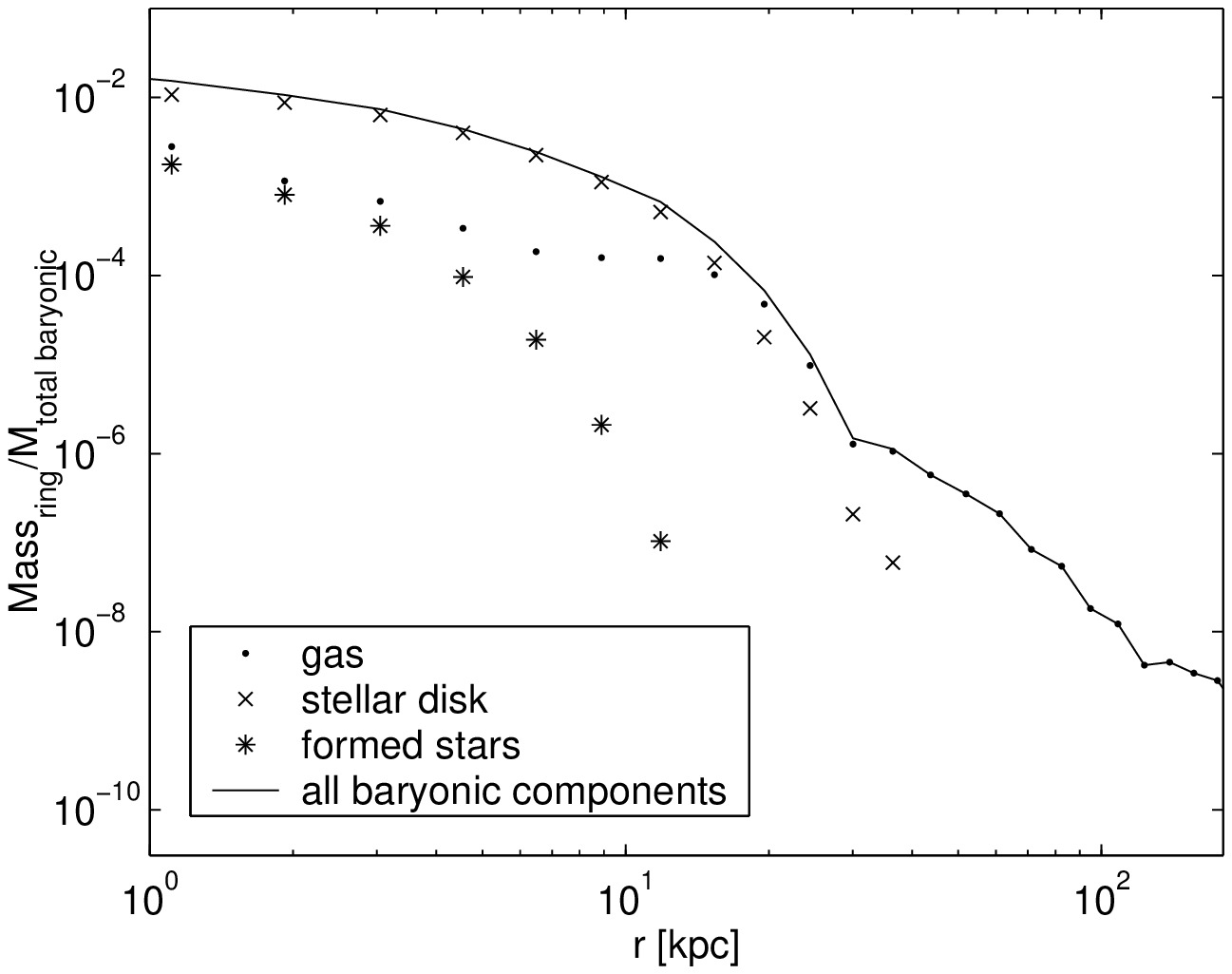}
\caption{Same as Fig. \ref{profile_AM} for model galaxy B}
\label{profile_BM}
\end{center}
\end{figure}

\begin{figure}[t]
\begin{center}
\includegraphics[width=8cm]{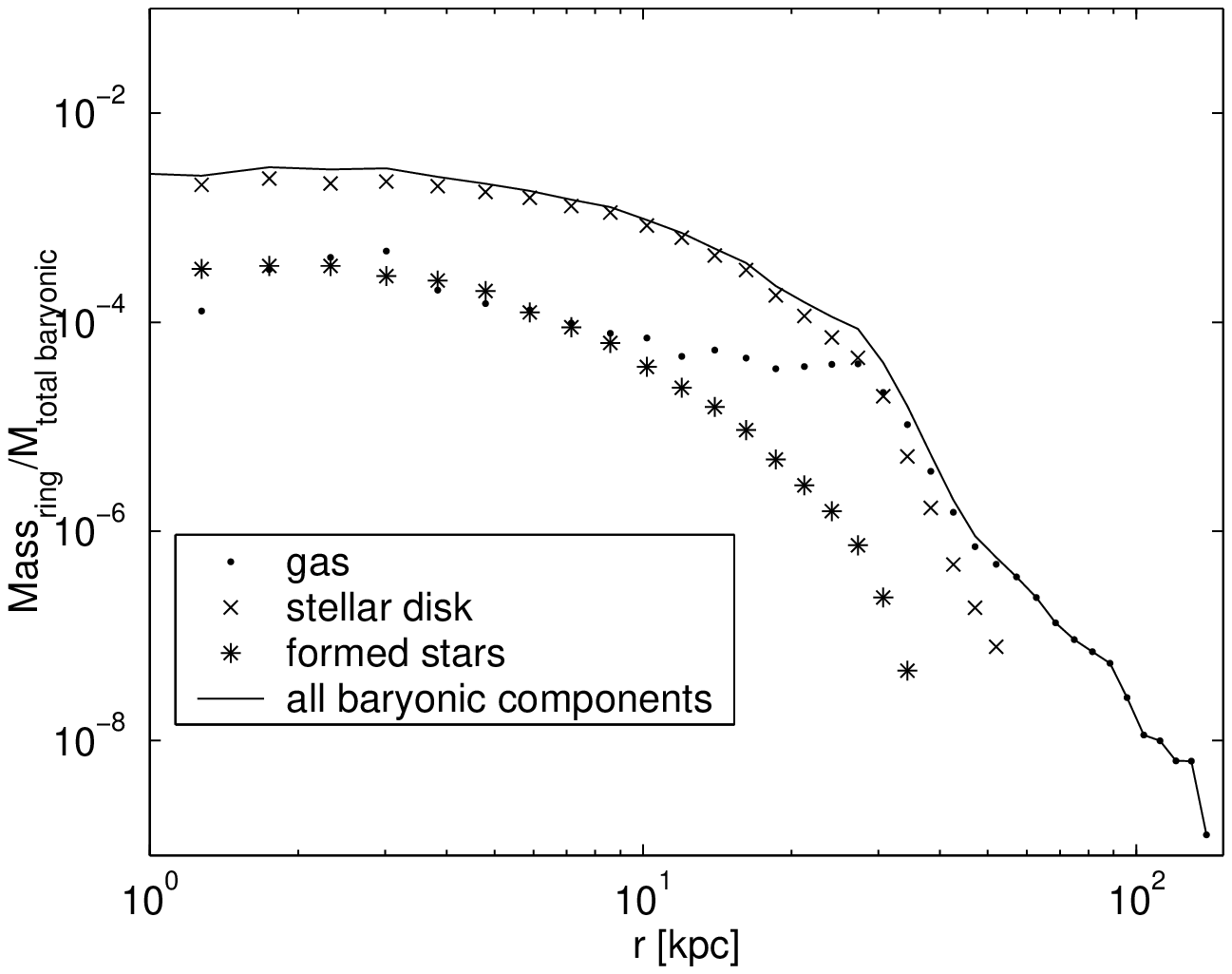}
\caption{Same as Fig. \ref{profile_AM} for model galaxy C}
\label{profile_CM}
\end{center}
\end{figure}

\begin{figure}[t]
\begin{center}
\includegraphics[width=8cm]{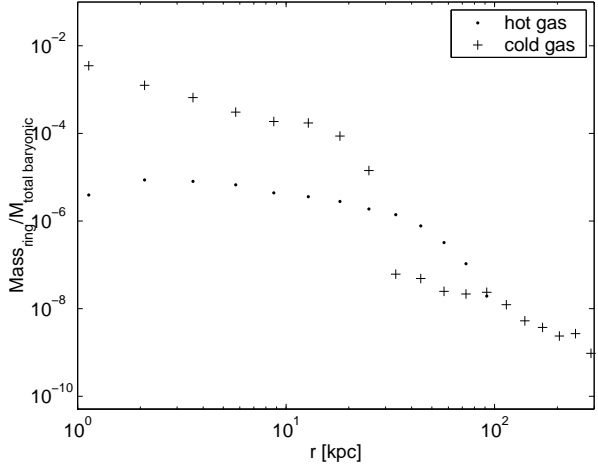}
\caption{Mass profiles of the hot and cold gas of the isolated
model galaxy B as a function of radius. The masses are given in
ratios of each component to the total baryonic mass of the whole
system. The binning is the same as in Fig. \ref{profile_AM}. The
profiles are given for the last timestep in our simulation (t=5
Gyr).} \label{profile_HCISM}
\end{center}
\end{figure}

\begin{figure*}[t]
\begin{center}
\includegraphics[width=16cm]{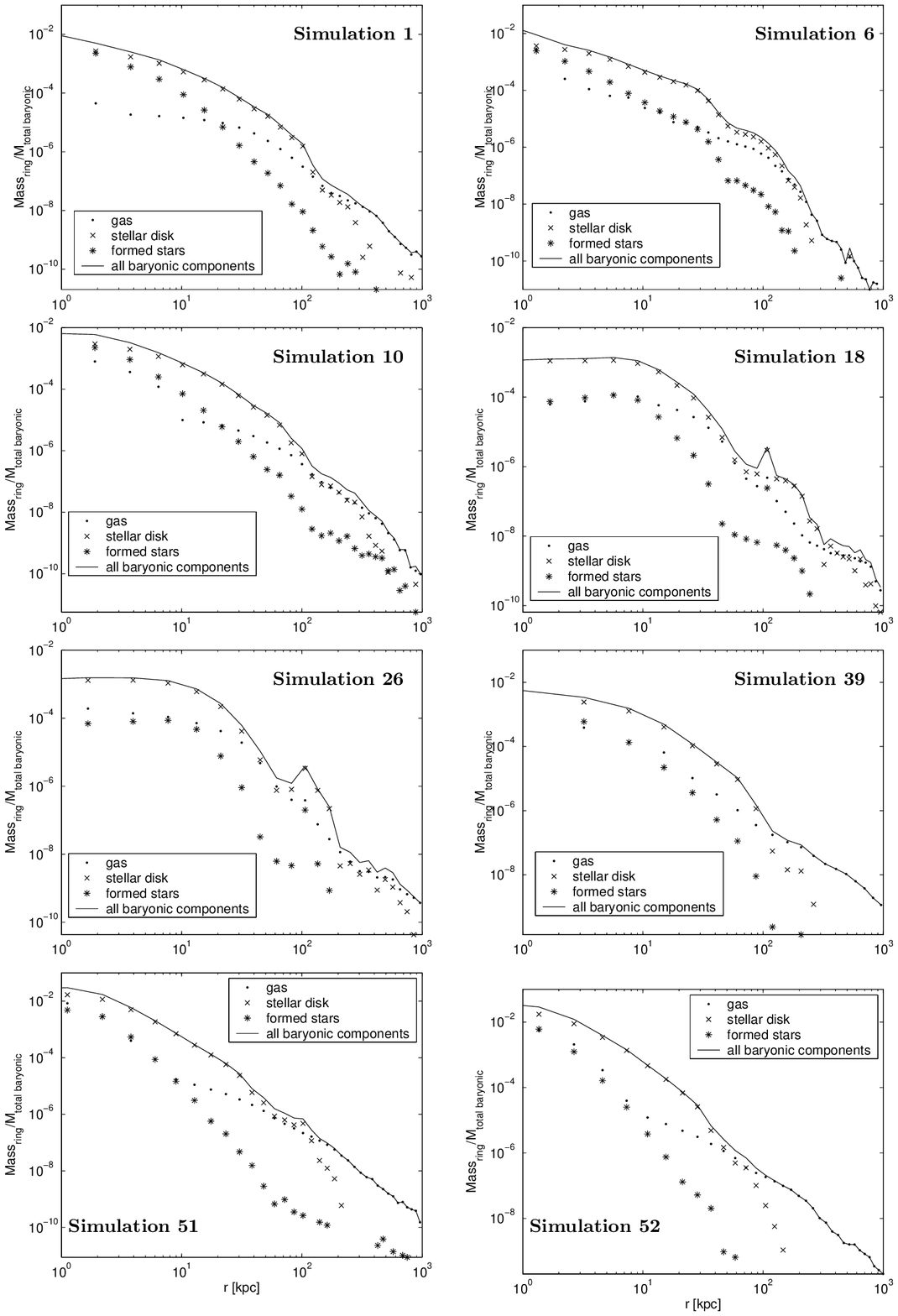}
\caption{Mass profiles of the baryonic components of the
interacting system as a function of the radius. The masses are
given in ratios of each component to the total baryonic mass of
the whole system. The binning geometry is given in Fig.
\ref{rings}. The center of the binning is the center of mass of
the baryonic components. The profiles are given for the last
timestep in our simulation (t=5 Gyr).} \label{profile1}
\end{center}
\end{figure*}

\noindent Figures \ref{profile_AM}-\ref{profile_CM} show the mass
profiles for the isolated systems A,B, and C at t=5 Gyr. In
comparison to the mass profiles of the interacting systems the
profiles are very flat in the innermost 3 kpc and they do not
reach as wide into the intergalactic space as the interacting
system (see Fig. \ref{profile1}). In the mass profiles of the
isolated model galaxies A and B there are kinks at 40 kpc and 30
kpc in the gas component, which coincide with the edges of the
stellar disks. This feature can be found in the interacting
systems as well (e.g. simulation 39). It marks the transition
between cold and hot gas (see
Fig. \ref{profile_HCISM}).\\
At first sight it is striking that in collisions A-A, A-C and B-B
(all equal mass mergers), the stars are finally more widely spread
in space than the gaseous matter components. The non-equal mass
mergers (simulations 17-28) on the other hand show the gaseous
matter to be almost always by a factor 2 more extended in space
than the stellar component. The maximum for the 80\% gaseous
matter cut off radius can be found in simulation 25 within 20.85
kpc. Figure \ref{profile1} gives the mass profiles for several
simulations. The masses are always given as ratios between mass of
the component or total baryonic mass in a ring with radius
r$_{i}$-r$_{i-1}$ and the total baryonic mass of the whole system.
Whereas simulations 1-16 (collisions A-A) do have a steeper
gradient in their mass profiles within a radius of about 10 kpc,
collisions between non-equal mass mergers (simulations 17-28) show
flat gradients in this range (see Fig. \ref{profile1}, simulations
1,6,10,18 and 26).

\begin{figure}[t]
\begin{center}
\includegraphics[width=8cm]{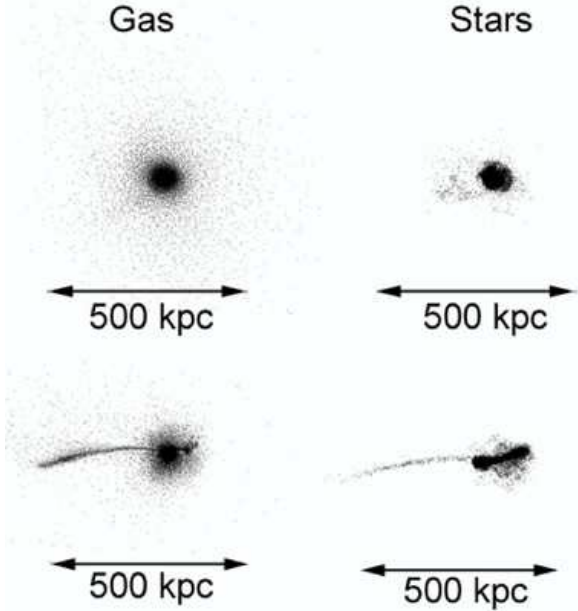}
\caption{Gas and mass distributions of simulation 13 (upper panel)
and simulation 25 (lower panel) at t=5 Gyr.}
\label{gas_mass_23_25}
\end{center}
\end{figure}

\noindent Simulations 18 and 23-27 show distinct baryonic mass
concentrations at r$\sim$100 kpc (see Fig. \ref{profile1},
simulations 18 and 26 for an example). In Fig.
\ref{gas_mass_23_25} the gaseous and stellar matter distributions
of simulations 13 and 25 are shown. Simulation 25 shows the
remnant of model galaxy B at about this distance of the centre of
mass. In addition a huge tail of gas and stars reaching nearly 500
kpc into the surrounding space can be seen. The corresponding mass
profiles, Fig. \ref{profile1} (simulations 18 and 26) show the
same features at large radii. Note that for r$>$200 kpc the
stellar component decreases dramatically in comparison to the
gaseous one. In contrast simulations 18, 22, 25, 26 and 27 show
nearly the same mass densities of gas and stellar matter up to 1
Mpc. If we compare the distinct mass concentrations at r$>$200 kpc
in simulations 18, 23, 24, 25, 26 and 27 there is a factor of
about ten more stellar than gaseous matter. It seems that the
passage of the smaller member (model galaxy B) has stripped off
nearly all of its gas. Additionally the gas of model galaxy A has
been vastly distributed, see Fig. \ref{gas_mass_23_25} lower
panel. The equal mass collisions (simulation 1-16) have common
trends in the spatial distribution of the different components. In
the innermost circles within radii $<$ 100 kpc the stellar matter
is the dominating component. At larger distances of the centre of
mass the gaseous matter becomes the dominating component. The
newly formed stars are more concentrated near the centre of
baryonic mass. The same behaviour can be found in simulations
29-44 (collisions between model galaxies A and C). Collisions
between model galaxies B show a similar distribution of baryonic
matter. In Fig. \ref{profile1} simulations 51 and 52 the gaseous
component dominates at radii above 100 kpc. The newly formed stars
are concentrated in a 10-20 kpc ring around the baryonic center.
In simulation 50 stars are formed even at higher radii than the
stellar disk reaches. At a distance of 200 kpc from the centre a
surface mass density of $\sim$1M$_{\sun}$ for newly formed stars
can be found (see Fig. \ref{profile1}).

\begin{table}
\caption[]{Radii containing 80\% of the mass of the baryonic
components. Col. 1: 80\% gas mass within radius [kpc] ;  Col. 1:
80\% stellar mass within radius [kpc] ; Col. 1: 80\% newly formed
stars within radius [kpc]}
\begin{tabular}{c c c c c }
\hline  \hline  & C 1 &  C2 & C3 \cr \hline galaxy A & 3.25 & 3.25
& 1.53 \cr galaxy B & 0.57 & 1.11 & 0.57 \cr galaxy C & 5.89 &
4.79 & 3.02 \cr \hline
\end{tabular}
\label{Cutoff radii}
\end{table}

\subsection{The difference between galactic winds and kinetic mass distribution}

\begin{figure}[h]
\begin{center}
\includegraphics[width=6cm]{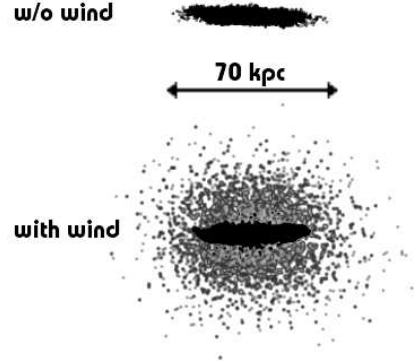}
\caption{Isolated galaxies after 5 Gyr of evolution with and
without wind routine switched on. Only the gas is shown edge on.}
\label{am_wind}
\end{center}
\end{figure}

In Fig. \ref{am_wind} the gas particles for the isolated model
galaxy A after 5 Gyr evolution are shown. The image displays the
galaxy edge-on in order to see the gas particles above and below
the disk due to galactic winds in the lower panel and the absence
of gas outside the disk in the upper panel in the case without. In
Table \ref{Cutoff radii} the cut off radii of the different
components, i.e. gaseous and stellar matter are given. A
comparison of the cut off radii of the different components
(gaseous matter, stars and newly formed stars) of the interacting
systems (see Tables \ref{big_table1} and \ref{big_table2}, Col.
18, 19 and 20) with the cut off radii of the isolated systems show
no common trend. E.g. the gaseous matter of simulations 1-16 is
not widely distributed in space, as in the isolated model galaxy
A. Only simulations 1,9 and 13 do not follow that trend and show
larger cut off radii. The reason is that the interacting systems
do convert much more gas into stars than the isolated ones.
Therefore over the simulation time more gas can be expelled by
galactic winds in the case of the isolated galaxy. It is important
to mention that the interaction between galaxies is a highly
dynamic process. At different simulation times of the interacting
systems the gaseous matter is distributed over the
intergalactic space in form of tidal tails and bridges.\\

\begin{figure}[h]
\begin{center}
\includegraphics[width=8.8cm]{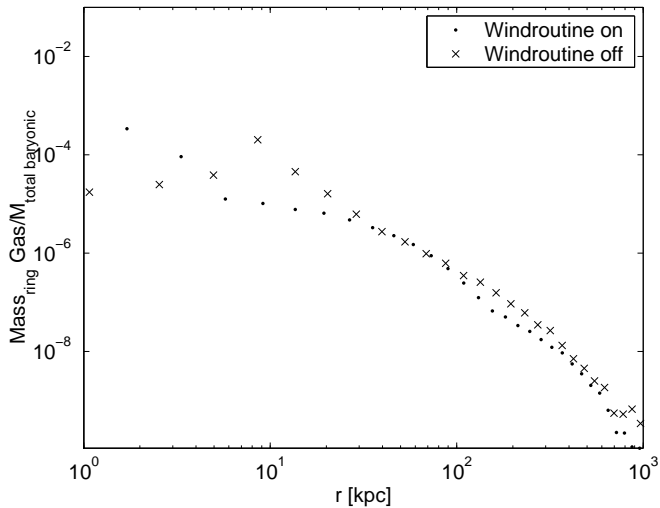}
\caption{Mass profiles for two simulations (Simulation 2, see
Table \ref{big_table1}), with wind routine switched on and off.}
\label{wind_on_off}
\end{center}
\end{figure}

\noindent As mentioned earlier, simulations 17-28 (interactions
between model galaxy A and B) show different results. Due to the
mass difference of the interacting systems (1:8) the gaseous
matter of the more massive interaction partner is thrown off by
the less massive galaxy as it travels through the gaseous disk of
its massive partner. A second mechanism is that model galaxy B
strips off almost all its gas as a consequence of the first
encounter with the other galaxie's disk.\\
To investigate the dependence of the mass profile on galactic
winds, we did galaxy collisions with and without galactic winds
switched on. One major result of our study is, that for our
mergers the -kinetic- spreading of gaseous matter is much more
efficient than the mass loss due to galactic winds. If the
galaxies were isolated, galactic winds can enrich several tens of
kpc of the surrounding intergalactic and intra-cluster medium very
efficient, see Fig. \ref{am_wind}. In Fig. \ref{wind_on_off} the
gas-mass profiles for one merger simulation with and without winds
are shown. In the outer parts of the merger, r $>$ 30 kpc, the
winds do not change the result. Therefore direct -kinetic-
spreading is the dominating process in the outer parts of the
merger remnant. Only in the inner parts the winds do change the
gas-mass profiles.

\clearpage

\section{Conclusions and discussion}

We present combined N-body/hydrodynamic simulations of interacting
spiral galaxies in order to investigate the overall star formation
and spatial distribution of baryonic matter due to different
encounter geometries and galaxy masses. We performed 56
simulations with average mass resolutions of $1.4\times 10^{5}$
M$_{\sun}$, $1.1\times 10^{6}$ M$_{\sun}$ and $9.5\times 10^{5}$
M$_{\sun}$ per particle. See Table
\ref{galaxyproperties_resolution} for details on the resolutions.
The main results of our simulations are:

\begin{enumerate}
\item The simulated interacting disk galaxies show an enhancement
of the ISFR of factors up to 5 and on average of a factor 2 in
comparison to isolated, undisturbed galaxies. This can be
explained in terms of duration and gas content of the interacting
system. As in Fig. \ref{ISFR} shown for some simulations the
duration of enhanced star formation is very short in comparison to
the quiescence star forming epochs. Of course this depends on the
relative velocities of the interacting systems. If the velocities
would be higher, like in centers of galaxy clusters, the
interacting times would be shorter and therefore the disturbances
lower and the ISFRs are not so enhanced. The result here holds
only, if the relative velocities for the interacting systems are
the same.

\item There is a strong dependence of the integrated star
formation rate (ISFR) on the minimum separation of the encounter.
The enhancement of the star formation scales linearly with a flat
slope ($\sim \!\!\! 1\!\times10^{-3}$ kpc$^{-1}$) after the first
encounter. If the minimum separation is larger than $\sim \! 5
\times$ the disk scale length R$_{d}$ at the first encounter the
second encounter does not provide an enhancement in the star
formation.This is due the fact, that the second encounter does not
occur within the simulation time. As the galaxies would have their
own SFR over longer periods the second encounter would not have as
much gas left as in the case of minimum separation less than
$5\times$R$_{d}$ encounters. Therefore the enhancement would be
lower. It is important to note that this result holds only for our
chosen parameters, therefore more investigation would be needed to
state a more global conclusion.

\item Counter-rotating interacting systems do not always have a
lower or higher ISFR in comparison to co-rotating systems. It
depends on the spatial alignment of the interacting galaxies. If
they interact edge-on, co-rotation does show a $\sim \!\! 10$\%
higher ISFR. On the other hand face-on collisions in general show
larger (on average 100\%) enhancements of the ISFR for the
counter-rotating case. This indicates that the duration of
interaction plays an important role, besides the geometry. In the
case of face-on interactions the first encounter is very short in
comparison to the edge-on interaction. Therefore the short
interaction times in the counter-rotating case does provide more
disturbances as the co-rotating case.

\item Collisions between model galaxy A (Milky Way type spiral
galaxy) and B (tiny spiral galaxy) do not show a significant
enhancement of the ISFR. On average 21\% more stars are formed
than in the isolated systems. As the mass ratio is 1:8 and the
disk scale length of model galaxy B is only half the size of model
galaxy A, those encounters do not provide as much disturbances in
the gaseous disks, as the collisions between model large model
galaxies. In addition the relative velocities of the smaller
members are higher as in the equal mass merger, which results in
shorter interaction times.

\item The mass profiles show that equal mass mergers do not
distribute stellar and gaseous matter as far as the non-equal mass
mergers. The cut off radii containing 80\% mass of the stellar and
gaseous components are the largest for collisions between model
galaxies A and B (mass ratio 8:1).
\end{enumerate}

\noindent Concluding we find that galaxy-galaxy interactions
enrich the intergalactic medium in two different processes:
indirectly by galactic winds and directly by redistributing
gaseous and stellar matter into huge volumes by interaction. If
the interacting galaxies have equal mass the galactic winds are
the dominating mechanism. But if the interacting system consists
of galaxies with a mass ratio of the order of 1:8, the direct
process is the dominating one. It is likely that the intergalactic
medium is highly enriched by this direct mechanism.

\section*{Acknowledgements}

The authors would like to thank Volker Springel for providing them
GADGET2 and his initial conditions generators. In addition the
authors are grateful to the anonymous referee for his/her
criticism that helped to improve the paper. The authors would like
to thank Giovanna Temporin, Wilfried Domainko and Magdalena Mair
for many useful discussions and Sabine Kreidl for corrections and
many useful suggestions. The authors acknowledge the Austrian
Science Foundation (FWF) through grant number P15868, the
UniInfrastrukturprogramm 2004 des bm:bwk Forschungsprojekt
Konsortium Hochleistungsrechnen and the bm:bwk Austrian Grid (Grid
Computing) Initiative and the Austrian Council for Research and
Technology Development.

\end{document}